\ifx\justbeingincluded\undefined
\documentclass{styles/svproc}
\usepackage[utf8]{inputenc}
\usepackage{latexsym}
\usepackage{amsmath}
\usepackage{amssymb}
\usepackage{mathrsfs}
\usepackage{graphicx}
\usepackage{epstopdf}
\usepackage{cite}
\usepackage{mathtools}
\usepackage{listings}
\usepackage{url}

\begin{document}
\mainmatter              
\fi

\title{Monolithic simulation of convection-coupled phase-change - verification and reproducibility}
\titlerunning{Monolithic simulation of convection-coupled phase-change}  
%
\author{Alexander G. Zimmerman \and Julia Kowalski}
\authorrunning{A. G. Zimmerman, J. Kowalski} 
%
\tocauthor{Alexander G. Zimmerman, Julia Kowalski}
\index{Zimmerman, A. G.}
\index{Kowalski, J.}
\institute{AICES Graduate School, RWTH Aachen University, 
	\newline Schinkelstr. 2, 52062 Aachen, Germany 
	\newline \email{zimmerman@aices.rwth-aachen.de}}

\maketitle              

\begin{abstract}

Phase interfaces in melting and solidification processes 
are strongly affected by the presence of convection in the liquid.
One way of modeling their transient evolution is 
to couple an incompressible flow model to an energy balance in enthalpy formulation.
Two strong nonlinearities arise, which account for
the viscosity variation between phases 
and the latent heat of fusion at the phase interface.

The resulting coupled system of PDE's can be solved by 
a single-domain semi-phase-field, 
variable viscosity, 
finite element method 
with monolithic system coupling 
and global Newton linearization \cite{danaila2014newton}.
A robust computational model for realistic phase-change regimes 
furthermore requires a flexible implementation based on sophisticated mesh adaptivity.
In this article, we present first steps towards implementing such a computational model
into a simulation tool which we call Phaseflow \cite{zimmerman2017phaseflow}.

Phaseflow utilizes the finite element software FEniCS \cite{alnaes2015fenics},
which includes a dual-weighted residual method 
for goal-oriented adaptive mesh refinement. 
Phaseflow is an open-source, dimension-independent implementation that, 
upon an appropriate parameter choice, reduces to classical benchmark situations 
including the lid-driven cavity and the Stefan problem. 
We present and discuss numerical results for these,
an octadecane PCM convection-coupled melting benchmark,
and a preliminary 3D convection-coupled melting example,
demonstrating the flexible implementation.
Though being preliminary, the latter is, to our knowledge, the first published 3D result for this method.
In our work, we especially emphasize reproducibility 
and provide an easy-to-use portable software container using Docker \cite{boettiger2015docker}.


\end{abstract}

\section*{Nomenclature}

\begin{tabular}{p{0.05\linewidth}p{0.45\linewidth}p{0.05\linewidth}p{0.45\linewidth}}

	$t$ & Time &
	
		$\gamma$ & Coefficient for penalty stabilization \\		
		
	$t_f$ & Final time &
	
		$h$ & Finite element cell diameter \\
			
	$\mathbf{x}$ & Spatial coordinates $\mathbf{x} = \begin{pmatrix} x & y & z \end{pmatrix}$ &
	
		$\mathbf{F}$ & The vector-valued strong form \\
	
	$x^*$ & phase-change interface's position &
		
		$F$ & Functional for the variational form \\ 
	
	$p$ & Pressure field &
		
		$\Omega$ & The spatial domain \\

	$\mathbf{u}$ & Velocity field &
	
		$\partial \Omega$ & Boundary of the spatial domain \\
	
	$T$ & Temperature field &
	
		$\partial\Omega_{T}$ & Dirichlet boundary for $T$ \\ 	
		
	$\mathbf{f}_B$ & Buoyancy force &
	
		$V$ & The scalar solution function space \\
	
	$r$ & Regularization smoothing factor &
	
		$\mathbf{V}$ & The vector solution function space \\
	
	$T_r$ & Regularization central temperature &
	
		$\mathbf{W}$ & Mixed finite element function space \\
	
	Pr & Prandtl number &
	
		$\mathbf{w}$ & System solution $\mathbf{w} = \begin{pmatrix} p & \mathbf{u} & T \end{pmatrix}$ \\
	
	Ra & Rayleigh number &

		$\boldsymbol{\psi}$ & Finite element basis functions \\

	Ste & Stefan number &
	
		$M$ & Adaptive goal functional \\
	
	$\mu$ & Dynamic viscosity &
	
		$\epsilon_M$ & Adaptive solver tolerance \\
	
	$\phi$ & Semi-phase-field &
	
		$\delta\mathbf{w}$ & Residual of linearized system \\
	
	$\mathrm{\Delta} t $ & Time step size &
	
		$\omega$ & Newton method relaxation factor \\
		
	$()_n$ & Values from discrete time $n$ &
		
		$()^k$ & Values from Newton iteration $k$ \\

\end{tabular}

\section{Introduction}
The melting and solidification of so-called phase-change materials (PCM's)
are relevant to many applications ranging from 
the design of latent heat based energy storage devices \cite{dutil2011review}, 
to ice-ocean coupling and its effects on Earth's climate \cite{dinniman2016modeling},
to the evolution of ocean worlds on the icy moons of our solar system \cite{hsu2015ongoing} 
and the design of robotic melting probes for their exploration \cite{kowalski2016navigation}.
The predictive modeling of phase-change systems is, however, challenging due to
1. strong nonlinearities at the phase-change interface (PCI),
2. the coupling of several physical processes (i.e multi-physics), and
3. a large range of relevant scales both in space and time (i.e. multi-scale).
Any mathematical model of a complex phase-change process 
hence manifests as a multi-scale and multi-parameter, nonlinear PDE system.
We aim to develop a robust and flexible model to simulate these systems.

\begin{figure}[tbp]	\begin{center}
	\includegraphics[height=5.5cm]{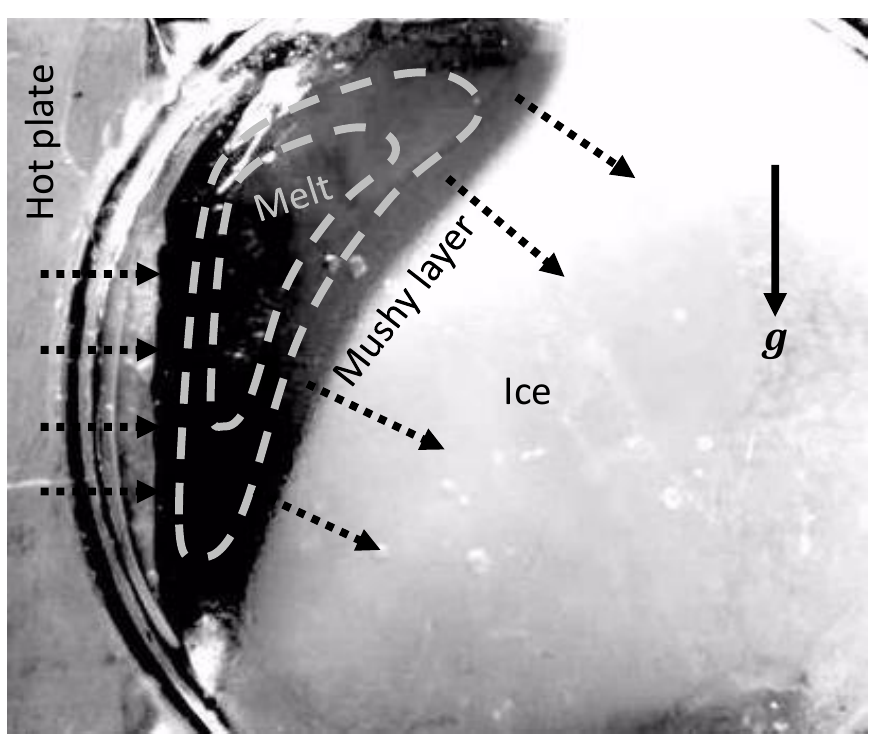}
	\caption[]
	{
		Sketch of the physical model superimposed on a melting experiment \cite{schueller2018preprint}.
		Heating from the left advances the phase-change interface (PCI) rightward. Buoyancy forces a convection cell, causing the upper PCI to advance more rapidly. In the absence of convection, the PCI would propagate as a planar front,
		indicating the significant impact of convection on the PCI's evolution.
	}
	\label{fig:PhysicalModel}
\end{center} \end{figure}

In this work, we will focus on phase-change in the presence of liquid convection.
Convection can have a tremendous effect on the evolution of phase-interfaces, as shown in Figure \ref{fig:PhysicalModel}.
A comprehensive introduction to melting and freezing without convection is given in \cite{alexiades1992mathematical}.
A mathematical model that accounts for convection-coupled phase-change is presented in \cite{batchelor2000perspectives}.
Therein mushy layer theory is introduced, which provides a model for understanding the PCI at macro-scale. 
The physical system is mathematically modeled by considering balance laws for mass, momentum, and energy,
with the momentum and energy balances coupled via buoyancy (which forces natural convection).

This constitutes a difficult multi-physics problem.
In a PCM domain,
certain physics dominate in the solid and liquid subdomains.
To solve the coupled problem, 
out of many approaches in the literature,
we employ the single-domain semi-phase-field enthalpy method.
With this approach,
the convection itself is handled by an incompressible Navier-Stokes flow model,
while the energy balance is modeled as the convection and diffusion of an enthalpy field.
The key idea is to solve the same equations on the entire domain.
This is commonly referred to as the \textit{fixed grid} approach,
for which various techniques are reviewed in \cite{voller1990fixed}.
Using an enthalpy method, phase-change latent heat effects are isolated 
to a source term in the energy balance \cite{voller1987enthalpy}.
Single-domain methods require a velocity correction scheme,
for which we select the variable viscosity method,
thereby treating the solid as a highly viscous fluid.

Efficiently applying this approach requires local mesh refinement to resolve the moving PCI.
This can be accomplished either with mesh refinement or front-tracking \cite{dutil2011review}.
Adopting the former approach,
we leverage the existing work on goal-oriented adaptive mesh refinement (AMR) methods, 
particularly the dual-weighted residual method \cite{bangerth2003adaptive}.
These days, goal-oriented AMR is widely practiced,
and multiple open-source software libraries provide this capability.

In the following, Section \ref{section:MathModel} describes the mathematical model.
Section \ref{section:NumericalMethods} presents numerical methods for the model's discretization and linearization.
Section \ref{section:Implementation} presents our implementation, Phaseflow \cite{zimmerman2017phaseflow},
based on the finite element software FEniCS \cite{alnaes2015fenics}.
Finally, Section \ref{section:Verification} presents verification via comparison to benchmark problems,
a convergence study for the 1D Stefan problem,
and a preliminary result for convection-coupled melting in a 3D domain.
We close with conclusions and an outlook.

\section{Mathematical Phase-Change Model} \label{section:MathModel}
\subsection{The Governing Equations} \label{section:GoverningEquations}
For the coupled phase-change system, out of many approaches from the literature, we adopt 
an enthalpy formulated \cite{voller1987enthalpy}, 
single-domain semi-phase-field \cite{belhamadia2004anisotropic},
variable viscosity model.
The mass and momentum balances (\ref{eq:mass}) and (\ref{eq:momentum}) are given by 
the incompressible Navier-Stokes equations,
with velocity field $\mathbf{u} = \mathbf{u}(\mathbf{x},t)$ 
and pressure field $p = p(\mathbf{x},T)$.
Invoking the Boussinesq approximation,
we extend the momentum equation with a temperature-dependent buoyancy forcing term $\mathbf{f}_B(T)$  
which couples it to the energy equation.
This approach is well established in the context of natural convection \cite{wang2010comprehensive}.
Furthermore, we consider the phase-dependent viscosity $\mu(\phi)$ in the momentum equation,
where $\phi = \phi(T)$ is the temperature-dependent phase.
For constant heat capacity, 
the energy balance in enthalpy form reduces to (\ref{eq:energy}), 
which is an extended form of the convection-diffusion equation
for the temperature field $T = T(\mathbf{x},t)$.
The diffusion term in (\ref{eq:energy}) is scaled by the Prandtl number $\mathrm{Pr}$.
The nonlinear source term $\frac{1}{\mathrm{Ste}} \frac{\partial}{\partial t} \phi$ 
accounts for the phase-change latent heat,
where $\mathrm{Ste}$ is the Stefan number.
Altogether the system of governing equations is
\begin{align} 
\label{eq:mass}
\nabla\cdot\mathbf{u} = 0\\
\label{eq:momentum}
\frac{\partial}{\partial t}\mathbf{u} + \left(\mathbf{u}\cdot\nabla\right)\mathbf{u}
+ \nabla p - \nabla\cdot\left(2 \mu(\phi) \mathbf{D(u)}\right) 
+ \mathbf{f}_B(T) = 0 \\
\label{eq:energy}
\frac{\partial}{\partial t} T 
- \frac{1}{\mathrm{Ste}} \frac{\partial}{\partial t} \phi
+ \nabla\cdot( T\mathbf{u})
- \frac{1}{\mathrm{Pr}}\mathrm{\Delta} T = 0
\end{align}
where the symmetric part of the rate-of-strain tensor is 
$\mathbf{D(u)} = \frac{1}{2}\left(\nabla\mathbf{u} + \left(\nabla\mathbf{u}\right)^\mathrm{T}\right)$.
The equations are unitless per the normalization in \cite{danaila2014newton},
shifted always such that $T = 0$ corresponds to the temperature of fusion.

\begin{figure}[tbp] \begin{center}
	\includegraphics[width=0.5\textwidth]{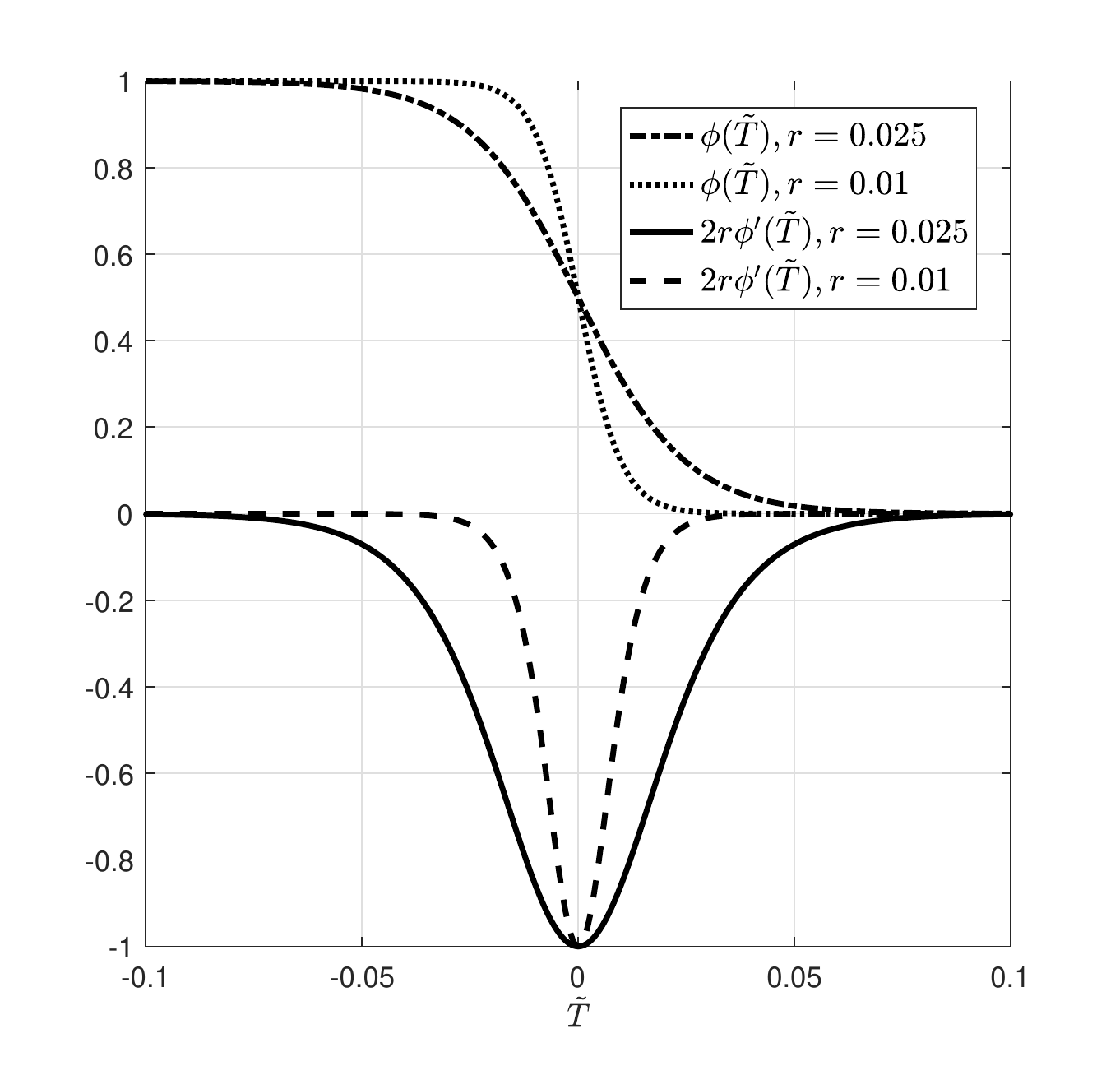}
	\caption{The regularized semi-phase-field (\ref{eq:phi}) with $T_r = 0$ and two values of $r$.
		For smaller $r$,
		$\phi(T)$ steepens
		and its derivative $\phi^{\prime}(T)$ approaches the Dirac delta function.
		Here the derivative is scaled by $2r$ for convenience.}
	\label{fig:Regularization}
\end{center} \end{figure}

We refer to $\phi$ as a \textit{semi}-phase-field \cite{belhamadia2004anisotropic},
because we do not treat $\phi$ as an additional unknown.
Rather, $\phi$ maps the temperature field to values between zero and one.
For this we use 
\begin{align} \label{eq:phi}
	\phi(T) &= \frac{1}{2}\left(1 + \tanh\frac{T_r - T}{r}\right),
\end{align}
where $T_r$ is the central temperature and $r$ is a smoothing parameter.
The Newton method requires the differentiability of (\ref{eq:phi}).
Subject to this requirement, there are many other useful regularizations from which to choose.
Figure \ref{fig:Regularization} plots $\phi$ and $\phi^{\prime}$.
For the Stefan problem in Section \ref{section:StefanProblem}, 
we simply set $T_r$ equal to the physical temperature of fusion
(i.e. $T_r = 0$ for the normalized system).
Following \cite{danaila2014newton}, for the coupled convection problem,
we set $T_r$ such that the strongest variation is localized near the newly appearing phase, 
e.g. liquid for melting.
Physically, the region where $0 < \phi < 1$ is analogous to a mushy layer.
In this sense, we can view $\phi$ as the solid volume-fraction.
Given $\phi$, we define the phase-dependent viscosity
as functions of the liquid and solid values 
(assuming constants $\mu_L$ and $\mu_S$ respectively) 
with
\begin{align} \label{eq:VariableViscosity}
	\mu(\phi) = \mu_L + (\mu_S - \mu_L)\phi,
\end{align}
With the variable viscosity method, the solid is treated as a highly viscous fluid, forcing its flow velocity to zero.
In general, not only viscosity varies in the PCM domain.
For example, the thermal material properties of water-ice 
vary significantly between the solid and liquid phases.
In the case of octadecane PCM's,
such as the benchmark in Section \ref{section:2DMeltingPCM},
we may treat these material properties as constants \cite{wang2010comprehensive}.

\subsection{The Initial Boundary Value Problem} \label{section:IBVP}

To simulate the time evolution of the PCM system,
we will solve (\ref{eq:strong}) as an initial boundary value problem 
subject to the initial values
\begin{align} \label{eq:InitialValues} \begin{split}
	\begin{pmatrix}
	p(\mathbf{x},t) , \mathbf{u}(\mathbf{x},t) , T(\mathbf{x}, t) 
	\end{pmatrix}^T &= 
	\begin{pmatrix} 
	0 , \mathbf{u}_0(\mathbf{x}) , T_0(\mathbf{x}) 
	\end{pmatrix}^T
	\quad \forall (\mathbf{x},t) \in \Omega \times 0
\end{split} \end{align}
and boundary conditions
\begin{align} \label{eq:BoundaryConditions} \begin{split}
	\mathbf{u(x},t) &= \mathbf{u}_D(\mathbf{x},t) \quad \forall (\mathbf{x},t) \in \partial\Omega \times ({0,t_f}], \\
	T(\mathbf{x},t) &= T_D(\mathbf{x},t) \quad \forall \mathbf{x} \in \partial\Omega_{T} \times ({0,t_f}], \\
	(\hat{\mathbf{n}}\cdot \nabla )T(\mathbf{x},t) &= 0 \quad \forall \mathbf{x} \in (\partial\Omega \setminus \partial\Omega_{T}) \times ({0,t_f}]
\end{split} \end{align}
For the applications in Section \ref{section:Verification}, 
we apply for all boundaries the velocity no-slip condition.
This implies homogeneous Dirichlet boundary conditions, $\mathbf{u}_D = \mathbf{0}$,
for all cases except the lid-driven cavity in Section \ref{section:LidDrivenCavity},
where the "moving" lid makes $\mathbf{u}_D$ non-homogeneous.
The notation $\partial \Omega_T$ refers to the boundary subdomain 
with non-homogeneous Dirichlet boundary conditions on the temperature.
Physically this models a thermal reservoir beyond the wall, 
keeping the wall at a constant temperature.
All other walls are prescribed homogeneous Neumann 
boundary conditions on the temperature,
which physically models thermally insulated (i.e. adiabatic) walls.

The generality of this initial boundary value problem
allows us to solve a variety of interesting benchmarks
by setting the appropriate parameters.
In Section \ref{section:Verification} we will demonstrate the 
2D lid-driven cavity, 
2D heat-driven cavity, 
1D Stefan problem,
2D convection-coupled melting of an octadecane PCM, 
and 3D convection-coupled melting.
Our implementation which we present in Section \ref{section:Implementation} is as versatile as the model.
Testing these benchmarks is accomplished with short Pythons scripts specifying the parameters.

\section{Numerical Methods} \label{section:NumericalMethods}

We base our work on the numerical approach from \cite{danaila2014newton}.
Therefore, we discretize in time via finite differences,
discretize in space via the finite element method (FEM) 
with a penalty formulation for stabilization,
couple the system monolithically,
and solve the nonlinear system globally via Newton's method.

\subsection{The Nonlinear Variational Problem} \label{section:NonlinearVariationalProblem}
We denote the system's solution as $\mathbf{w} = \begin{pmatrix} p & \mathbf{u} & T \end{pmatrix}$
and  write the strong form (\ref{eq:mass}), (\ref{eq:momentum}), and (\ref{eq:energy}) 
as the vector-valued functional
\begin{equation} \label{eq:strong}
	\mathbf{F}(\mathbf{w}) = \mathbf{0}
	\quad \forall (\mathbf{x},t) \in \Omega \times ({0,t_f}],
\end{equation}
with the three components respectively being the mass, momentum, and energy equations.
To stabilize the finite element method,
following the penalty formulation in \cite{danaila2014newton}, 
we add a pressure stabilizing term $\gamma p$ to the mass component.
The coefficient $\gamma$ is generally small, and in this case is $\gamma = 10^{-7}$.
We consider 1D, 2D, and 3D spatial domains,
respectively allowing for $\Omega \subset \mathbb{R}^1$, $\Omega \subset \mathbb{R}^2$, or $\Omega \subset \mathbb{R}^3$.

Per the standard Galerkin finite element method,
we write the variational (i.e. weak) problem whose solution approximates (\ref{eq:strong}).
To obtain the monolithic system, we use mixed finite elements \cite{brezzi1991mixed}.
Therefore, we multiply (\ref{eq:strong}) from the left 
by test functions $\boldsymbol{\psi} = \begin{pmatrix} \psi_p & \boldsymbol{\psi}_u & \psi_T \end{pmatrix}$
and integrate by parts over the domain.
Finally, we employ the fully implicit Euler method
by substituting the time discretizations
$\partial_t \mathbf{u} = \mathrm{\Delta} t ^{-1}(\mathbf{u}_{n+1} - \mathbf{u}_n),
\partial_t T = \mathrm{\Delta} t ^{-1}(T_{n+1} - T_n),
\partial_t \phi = \mathrm{\Delta} t ^{-1}(\phi_{n+1} - \phi_n)$.

Altogether, for homogeneous Neumann boundary conditions,
this yields the time-discrete nonlinear variational form
\begin{equation}\begin{split} \label{eq:VariationalForm}
	F(\boldsymbol{\psi},\mathbf{w}) = \int_{\Omega} 
	\begin{pmatrix}
		\psi_p & \boldsymbol{\psi}_u & \psi_{T}
	\end{pmatrix}
	\mathbf{F}(p,\mathbf{u},T) d\mathbf{x} = \\
	b(\mathbf{u},\psi_p) - (\psi_p,\gamma p)  \\
	+ 
	\left(
		\boldsymbol{\psi}_u,
		\frac{1}{\mathrm{\Delta} t}(\mathbf{u}_{n+1} - \mathbf{u}_n)
		+ \mathbf{f}_B(T)
	\right) 
	+ c(\mathbf{u};\mathbf{u},\boldsymbol{\psi}_u)
	+ b(\boldsymbol{\psi}_u,p)
	+ a(\mu(\phi);\mathbf{u},\boldsymbol{\psi}_u) \\
	+\frac{1}{\mathrm{\Delta} t} 
	\left( 
		\psi_{T},
		T_{n+1} - T_n 
		- \frac{1}{\mathrm{Ste}}
		\left(
			\phi(T_{n+1}) - \phi(T_n)
		\right)
	\right) 
	+ 
	\left(
		 \nabla \psi_{T}, 
		 \frac{1}{\mathrm{Pr}}\nabla T
		 - T\mathbf{u}
	\right)
\end{split}\end{equation}
where we use the short-hand 
$(u,v) = \int_{\Omega}uv d\mathbf{x}$, $(\mathbf{u},\mathbf{v}) = \int_{\Omega}\mathbf{u}\cdot\mathbf{v} d\mathbf{x}$ 
for integrating inner products.
Additionally, we use a common notation \cite{donea2003finite} 
for the linear, bilinear, and trilinear forms 
of the variational Navier-Stokes equations 
\begin{align}\begin{split}
&a: \mathbf{V}\times\mathbf{V} \rightarrow \mathbb{R}, 
\quad a(\mu;\mathbf{u},\mathbf{v}) = 2\int_{\Omega} \mu\mathbf{D(u)}:\mathbf{D(v)} d\mathbf{x} \\
&b: \mathbf{V}\times V \rightarrow \mathbb{R}, 
\quad b(\mathbf{u},p) = - \int_{\Omega} p \nabla \cdot \mathbf{u} d\mathbf{x} \\
&c: \mathbf{V}\times\mathbf{V}\times\mathbf{V} \rightarrow \mathbb{R},
\quad c(\mathbf{u}; \mathbf{z}, \mathbf{v}) = \int_{\Omega} \mathbf{v}^{\mathrm{T}} \left(\nabla \mathbf{z}\right) \mathbf{u} d \mathbf{x}
\end{split}\end{align}
Given (\ref{eq:VariationalForm}), we write the variational problem as

Find $\mathbf{w} \in \mathbf{W} $ such that 
\begin{equation} \label{eq:VariationalProblem}
	F(\boldsymbol{\psi},\mathbf{w}) = 0
	\quad \forall \boldsymbol{\psi} \in \hat{\mathbf{W}}
\end{equation}
where $\mathbf{W} = V \times \mathbf{V} \times V$ and $\hat{\mathbf{W}} = \hat{V} \times \hat{\mathbf{V}} \times  \hat{V}$.
This distinction comes from how non-homogeneous boundary conditions are often handled in finite element method implementations.
The solution is split into homogeneous and non-homogeneous parts, 
the former is found, and then the latter is reconstructed.
Therefore, for example, $T$ belongs to the space
$V = \left\{v \in H^1_0(\Omega): v = T_D \ \mathrm{on} \ \partial\Omega\right\}$,
while $\psi_T$ belongs to the space
$\hat{V} = \left\{v \in H^1_0(\Omega): v = 0 \ \mathrm{on} \ \partial\Omega \right\}$,
where $H^1_0(\Omega)$ is the classical Hilbert space fulfilling requirements for continuity and compact support.

The vector-valued function space $\mathbf{V}$ depends on the spatial domain's dimensionality,
e.g. $\mathbf{V} = V \times V$ in 2D.
For the incompressible Navier-Stokes solution,
we use the Taylor-Hood element \cite{donea2003finite},
i.e. we restrict the pressure solution to piece-wise continuous linear Lagrange polynomials,
and we restrict the velocity solution to piece-wise continuous quadratic Lagrange polynomials.
We use the same polynomial space for temperature and pressure.

\subsection{Linearization via Newton's Method}
We apply Newton's method by solving a sequence (indexed by superscript $k$) of linear problems 

Find $\mathbf{\delta\mathbf{w}} \in \mathbf{W}$ such that
\begin{align} \label{eq:LinearizedProblem}
	D_{\mathbf{w}} F(\boldsymbol{\psi},\mathbf{w}^k;\delta \mathbf{w}) = F(\boldsymbol{\psi}, \mathbf{w}^k)
	\quad \forall \boldsymbol{\psi} \hat{\mathbf{W}}
\end{align}
for the residual $\delta \mathbf{w}$ which updates the solution $\mathbf{w}^k$,
converging $\mathbf{w}^k$ to the approximate solution of the nonlinear problem.
The Gâteaux derivative of $F(\boldsymbol{\psi},\mathbf{w}^k)$,
defined as
$D_{\mathbf{w}}F(\boldsymbol{\psi},\mathbf{w}^k;\delta \mathbf{w}) \equiv
\frac{d}{d \epsilon} F(\boldsymbol{\psi},\mathbf{w}^k + \epsilon \delta \mathbf{w})|_{\epsilon=0}$,
is given by
\begin{equation}\begin{split} \label{eq:GateauxDerivative}
D_{\mathbf{w}}F(\boldsymbol{\psi},\mathbf{w}^k;\delta \mathbf{w}) =
b(\delta \mathbf{u},\psi_p) - (\psi_p, \gamma \delta p)\\
+ \left(\boldsymbol{\psi}_u,\frac{1}{\mathrm{\Delta} t}\delta \mathbf{u} + \delta T \mathbf{f}_B^{\prime}(T^k) \right)
+ c(\mathbf{u}^k;\delta\mathbf{u},\boldsymbol{\psi}_u) 
+ c(\delta\mathbf{u};\mathbf{u}^k,\boldsymbol{\psi}_u) \\
+ b(\boldsymbol{\psi}_u,\delta p)
+ a\left(\delta T\mu^{\prime}(T^k);\mathbf{u}^k,\boldsymbol{\psi}_u \right) 
+ a\left(\mu(T^k);\delta\mathbf{u},\boldsymbol{\psi}_u\right)
\\
+ 
\frac{1}{\mathrm{\Delta} t}
\left(
	\psi_{T} \delta T,
	1 - \frac{1}{\mathrm{Ste}} \phi^{\prime}(T^k)
\right)
+ 
\left(
	\nabla \psi_T,
	\frac{1}{\mathrm{Pr}} \nabla \delta T
	- T^k \delta \mathbf{u}
	- \delta T \mathbf{u}^k
\right),
\end{split}\end{equation}

For a given $\mathbf{w}^k$ (\ref{eq:GateauxDerivative}) and hence (\ref{eq:LinearizedProblem}) 
are linear with respect to the unknown Newton residual $\delta \mathbf{w}$.
In typical fashion, we define the test functions,
solution and Newton residual 
as linear combinations of the same basis.
Upon selecting a mesh and concrete basis,
this allows (\ref{eq:LinearizedProblem}) to be re-written
as a discrete linear system of the form $\mathbf{A} \mathbf{x} = \mathbf{b}$,
which can be efficiently solved on a computer by standard methods and software.
In this work, we directly solve each linear system with LU decomposition 
(using the interface of FEniCS, discussed in Section \ref{section:Implementation}).

We use the latest discrete time solution as the initial guess for the Newton solver, 
i.e. we initialize $\mathbf{w}^0 = \mathbf{w}_n$.
After each iteration solving for $\delta \mathbf{w}$,
we update the solution with the relaxed residual 
$\mathbf{w}^{k + 1} \coloneqq \mathbf{w}^k + \omega \delta \mathbf{w}$,
where $0 < \omega \le 1$ is a relaxation factor.
Relaxing Newton's method is often useful for highly nonlinear problems.
The 1D and 2D results in Section \ref{section:Verification} use the full Newton ($\omega = 1$) method.
So far only the preliminary 3D result requires relaxation, where we will set $\omega = 0.8$.

\subsection{Adaptive Mesh Refinement}
The single-domain approach requires local mesh refinement,
or else the computational cost would quickly become impractical, especially in 3D.
Furthermore, we cannot \textit{a priori} prescribe where to locally refine the mesh,
since our goal is to predict the position of the PCI.
This means we must employ adaptive mesh refinement, now commonly referred to as AMR.
The theory of AMR fundamentally requires an error estimator,
which exist for many interesting problems \cite{kelly1983posteriori}.
In the context of phase-change simulations,
hierarchical error estimators have been derived for the Stefan problem \cite{belhamadia2004anisotropic};
but no such rigorous work has been completed for the problem with coupled convection.
Promising results are reported in \cite{danaila2014newton},
wherein they used a mesh adaptivity procedure by metric control 
which was particular to the software library and its Delaunay mesh generation procedure.
Unfortunately, it is unclear exactly which metrics were used for adaptivity in those results.

We instead employ goal-oriented AMR \cite{bangerth2003adaptive}.
To discuss AMR, let us briefly write the spatially discrete problem 
which is dependent on a mesh $\mathbf{W}_h \subset \mathbf{W}(\Omega)$:

Find $\mathbf{w}_h \in \mathbf{W}_h \subset \mathbf{W}(\Omega)$ such that
\begin{align} \label{eq:DiscreteNonlinear}
	F(\boldsymbol{\psi}_h,\mathbf{w}_h) = 0 \quad \forall \boldsymbol{\psi}_h \in \hat{\mathbf{W}}_h \subset \hat{\mathbf{W}}
\end{align}
Goal-oriented AMR requires some goal functional $M(\mathbf{w})$
to be integrated over the domain $\Omega$. 
The goal-oriented adaptive solution of (\ref{eq:DiscreteNonlinear}) can then be written:

Find $\mathbf{W}_h \subset \mathbf{W}(\Omega)$ and $\mathbf{w}_h \in \mathbf{W}_h$ such that
\begin{align}
	\left|M(\mathbf{w}) - M(\mathbf{w}_h) \right| < \epsilon_M
\end{align}
where $\epsilon_M$ is some prescribed tolerance.
Since $M(\mathbf{w})$ is unknown, we still require an error estimator.
To this end, we use the dual-weighted residual method, as implemented in FEniCS \cite{alnaes2015fenics}.
Computing cell-wise error estimates requires 
solving a linearized adjoint (with respect to the goal) problem.
The primal and adjoint problems are solved on a hierarchy of meshes.
Computing the linearized adjoint solution on each mesh is relatively cheap 
compared to solving the nonlinear primal problem.
See \cite{bangerth2003adaptive} for a full explanation of the method.
For the adaptive solutions in this work, we set the goal functional
\begin{equation} \label{eq:AdaptiveGoalPhase}
	M(\mathbf{w}) = \int_{\Omega} \phi(T) d\mathbf{x}
\end{equation}
which represents the volume of solid material remaining in the domain.

\section{Implementation} \label{section:Implementation}
The method presented in this paper, along with its application to a series of test problems,
were implemented by the author 
into an open-source Python module named Phaseflow \cite{zimmerman2017phaseflow},
using the finite element library FEniCS \cite{alnaes2015fenics}.

\subsection{FEniCS}
The abstract interface of FEniCS \cite{alnaes2015fenics} makes it an ideal library
to quickly prototype models and methods
which use FEM.
Additionally, standard algorithms for
Newton linearization and goal-oriented AMR
are already implemented.
FEniCS is an umbrella project.
A major back-end component is the C++ library DOLFIN, 
which implements most of the classes and methods seen by the user.
Python interfaces are generated mostly automatically,
and indeed FEniCS is primarily used as a Python module.
From a conceptual perspective, we are most interested in two components of FEniCS:
the Unified Form Lanugage (UFL) and the FEniCS Form Compiler (FFC).
UFL allows us to write down the abstract discrete variational (i.e. weak) form
in a way that is understood by FEniCS.
This means that to implement variational forms, we write them as source code almost character-for-character.
The FEniCS Form Compiler (FFC) then automatically implements the FEM matrix assembly routine 
in optimized C++ code using just-in-time (JIT) compiling.
 
\subsection{Phaseflow}
Phaseflow \cite{zimmerman2017phaseflow},
created by the author,
is a Python module maintained openly on GitHub.
Phaseflow implements the methods of this paper
using the open-source finite element library FEniCS \cite{alnaes2015fenics}.
The interface allows users to run these methods for the variational form (\ref{eq:VariationalForm})
with any set of similarity parameters $\mathrm{Ste}, \mathrm{Pr}$,
buoyancy model $\mathbf{f}_B(T)$,
regularization $\phi(T)$,
liquid and solid viscosities $\mu_L$ and $\mu_S$,
time step size $\mathrm{\Delta} t$,
and stabilization coefficient $\gamma$.
The initial values (\ref{eq:InitialValues}) are set with a function in the solution space.
Combined with the built-in FEniCS method, this allows the user to 
interpolate general mathematical expressions written in the C syntax,
or to use an existing solution.
Dirichlet boundary conditions (\ref{eq:BoundaryConditions})
are also set with general C mathematical expressions.
Additionally the interface allows users to control algorithm parameters,
such as tolerances for the solvers, 
the Newton relaxation parameter $\omega$, 
and other options.

The interface accepts any FEniCS mesh object,
which can either be generated from scratch or can be converted from a standard format via the FEniCS library.
Leveraging the abstract interfaces of FEniCS, Phaseflow is dimension-independent, and has been applied to 1D, 2D, and 3D spatial domains.
In this work, unit square and rectangular prism domains have been used.
Phaseflow writes all solutions to disk in the efficient XDMF+HDF5 format 
using the built-in FEniCS interface.
Furthermore, this is combined with the H5Py library 
to write checkpoint files 
for easily restarting simulations at later times.
Phaseflow has other ease-of-use features,
such as a stopping criterion for steady state solutions.
This was useful for the heat-driven cavity test in Section \ref{section:HeatDrivenCavity},
where the Newton method required a small time step size.

\subsection{Reproducibility with Docker}
Phaseflow leverages open technology, and it is meant to contribute further to open scientific and engineering research.
Open research in the computational sciences is often hindered by the difficulty of reproducing results,
primarily because of the ever-increasing complexity of computing environments \cite{boettiger2015docker}.
Recently, the software Docker has emerged 
as a technology which greatly facilitates reproducibility in the computational sciences. 

Built upon an official FEniCS Docker image, we provide a Phaseflow Docker with all dependencies.
Enabled by the Docker image and Phaseflow's test suite,
changes to the master branch on GitHub 
are continuously tested using the Travis-CI continuous integration service \cite{boettiger2015docker}.
Docker has been central to the development of Phaseflow. 
Initially, a primary motivation for its use 
was the existence of Docker images where most of the dependencies of Phaseflow 
are already installed and maintained by the FEniCS developers.
This allows us to spend more time implementing models,
and less time building tool chains.
From there, maintaining a Docker image which runs Phaseflow takes little effort.
With the Phaseflow Docker image in hand,
we were then able to set up a continuous integration process with remarkably little effort.

The vast majority of Phaseflow's source code is covered with a suite of unit tests and integration tests written with PyTest.
Most results in this paper are included in Phaseflow's test suite,
while the others are covered by the example scripts and notebooks in its repository \cite{zimmerman2017phaseflow}.
Such results are easy to reproduce,
and we highly encourage the reader to do this.
To further aid reproducibility,
a specific version of Phaseflow has been archived (see the release versions at \cite{zimmerman2017phaseflow}) to coincide with this publication.

\section{Verification} \label{section:Verification}

To verify our implementation,
we consider a series of examples, including
the lid-driven cavity benchmark from \cite{ghia1982high} extended with a solid subdomain,
the heat-driven cavity benchmark from \cite{wang2010comprehensive},
an approximation of the analytical 1D Stefan problem from \cite{alexiades1992mathematical} with and without AMR,
a preliminary 2D convection-coupled melting demonstration with AMR,
and a preliminary 3D convection-coupled melting demonstration without AMR.
Furthermore, for the Stefan problem,
we verify the convergence orders of the temporal and spatial discretizations, and of the Newton method.
Phaseflow's flexible interface described in Section \ref{section:Implementation} allows us to implement
each of these applications with minimal effort,
and to test the same lines of source code, increasing our confidence with the implementation.

\subsection{Lid-Driven Cavity with Solid Subdomain} \label{section:LidDrivenCavity}
To test phase-dependent viscosity,
we consider an extension of the lid-driven cavity benchmark
based on data published in \cite{ghia1982high}.
The standard benchmark uses a unit square geometry.
We extend the geometry below the bottom wall,
and set the temperature in the new region below the freezing temperature,
such that the liquid subdomain per (\ref{eq:phi}) covers the original unit square geometry.
Therefore the initial temperature values are
\begin{equation}
	T_0(\mathbf{x}) = 
	\begin{cases}
		T_c & $for $ y \leq 0, \\
		T_h & $otherwise$
	\end{cases}
\end{equation}
where we chose $T_h = 1$ and $T_c = -1$.
For the semi-phase-field (\ref{eq:phi}) 
we set central temperature $T_r = -0.01$ and smoothing parameter $r = 0.01$.
We set variable viscosity (\ref{eq:VariableViscosity}) with $\mu_L = 1$ and $\mu_S = 10^8$.
To capture this strong variation,
we refined all cells of the initial mesh which touched coordinate $y = 0$ for four refinement cycles.
Note that in this case, the local refinement is not adaptive.
This problem at steady state fits our general model (\ref{eq:VariationalForm}) with
null buoyancy $\mathbf{f}_B = \mathbf{0}$
and with an arbitrarily large Prandtl number to nullify thermal conduction.

\begin{figure}[tbp]\begin{center}
		\includegraphics[height=5.5cm]{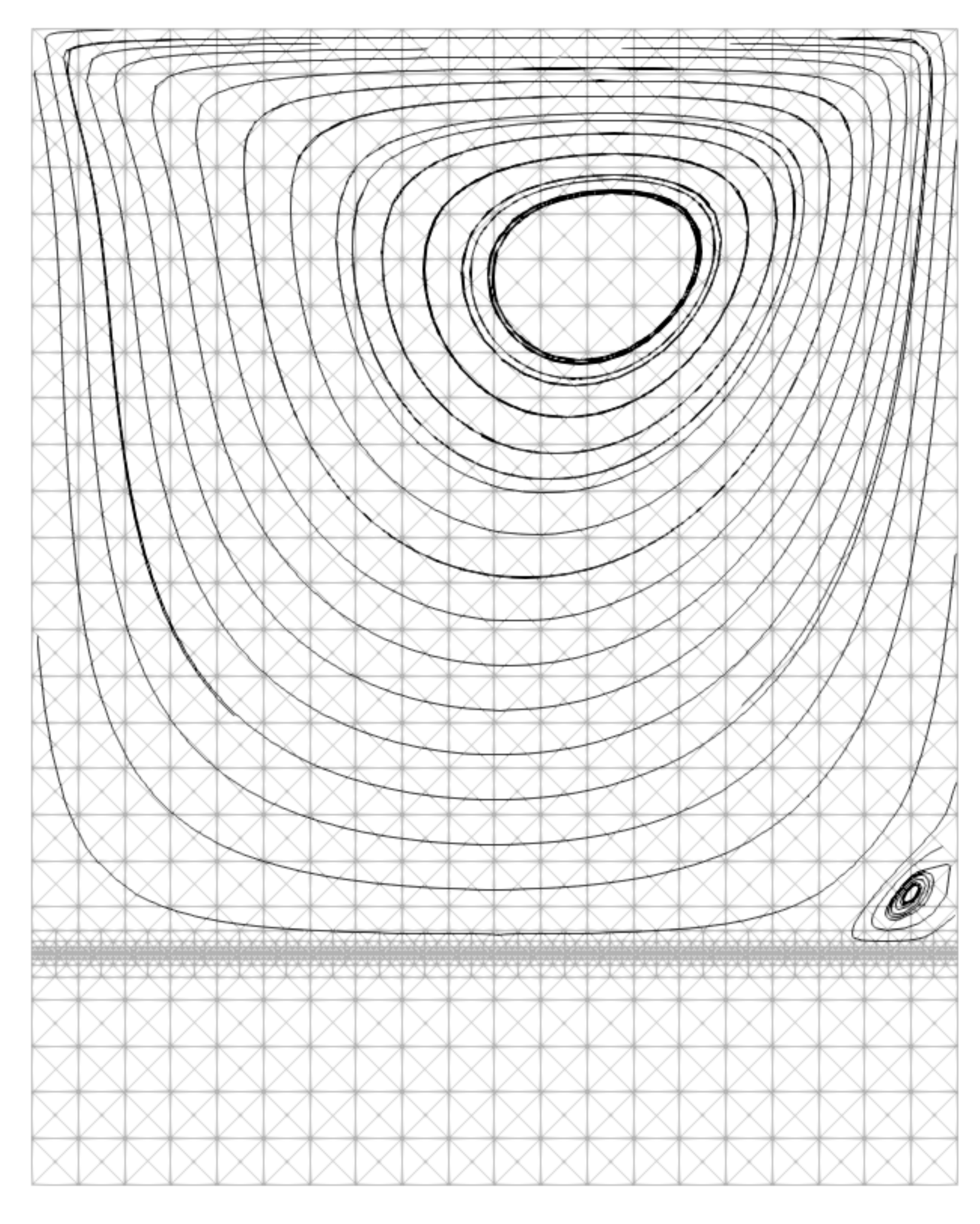}
		\includegraphics[height=5.5cm]{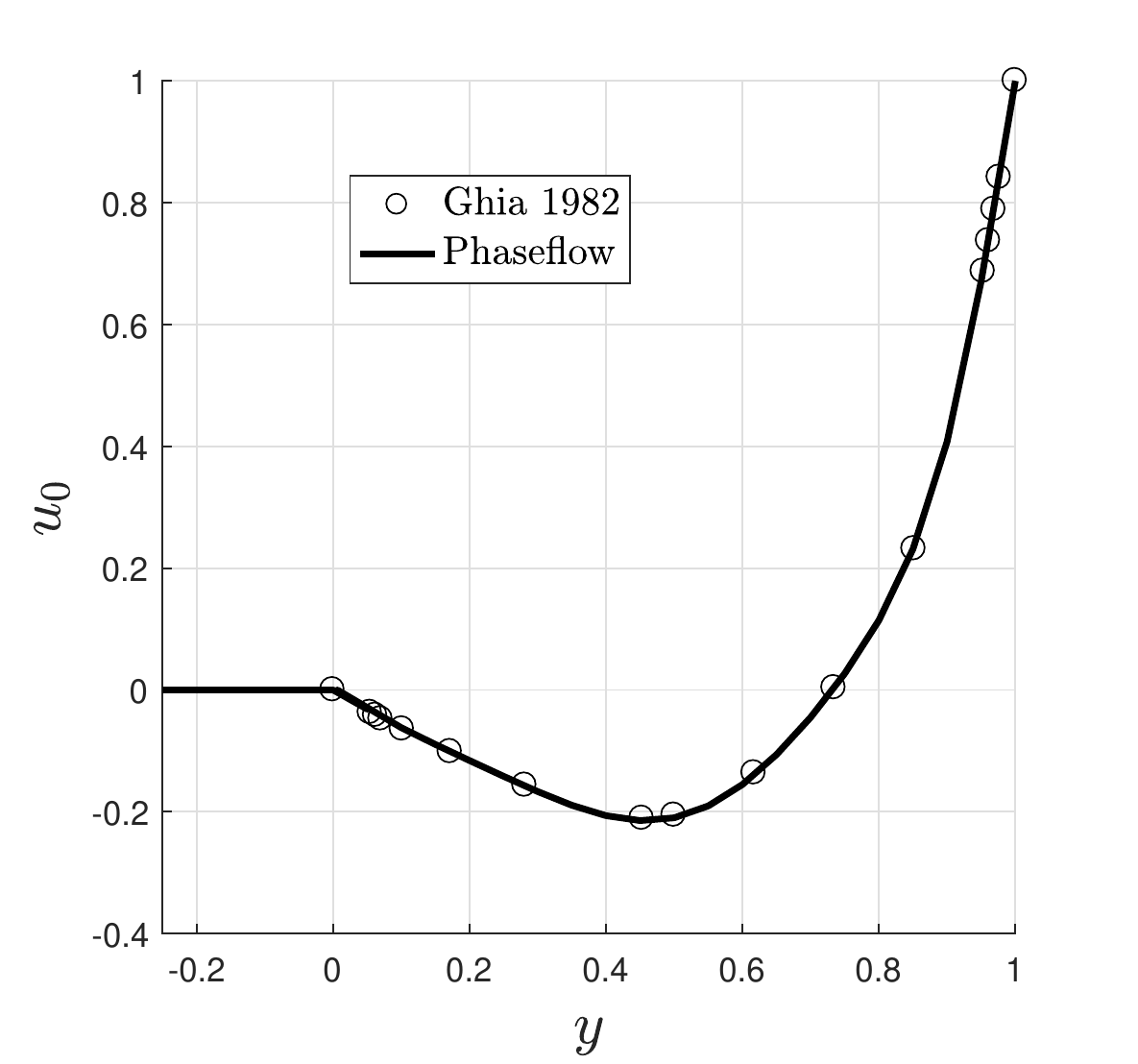}
		\caption{Left: Result from lid-driven cavity test ($\mathrm{Re} = 100$).
			The mesh (which is locally refined near the PCI) is shown in translucent gray.
			Velocity streamlines are shown in black.
			As expected, the moving lid causes circulation in the cavity,
			and we see the expected recirculation zone in the bottom-right corner of the liquid subdomain.
			\newline
			Right: Horizontal velocity sampled from the vertical centerline,
			compared to benchmark data published in \cite{ghia1982high}.}
		\label{fig:LidDrivenCavity}
\end{center} \end{figure}

As boundary conditions $\mathbf{u}_D$, we set positive horizontal velocity on the lid,
zero horizontal velocity away from the lid,
and zero vertical velocity everywhere.
We set initial velocity $\mathbf{u}_0$ similarly,
which serves as a suitable initialization for solving until steady state with Newton's method.
Because Phaseflow implements only the unsteady problem,
we solve a single large time step
to obtain the approximate steady state solution.
It is advantageous to model the steady problem this way,
when the goal is to verify components of our unsteady implementation.
Figure \ref{fig:LidDrivenCavity} shows and discusses the result.
The solution agrees very well with the published benchmark data.

\subsection{Heat-Driven Cavity} \label{section:HeatDrivenCavity}
\begin{figure}[tbp]
\begin{center}
	\includegraphics[height=5.5cm]{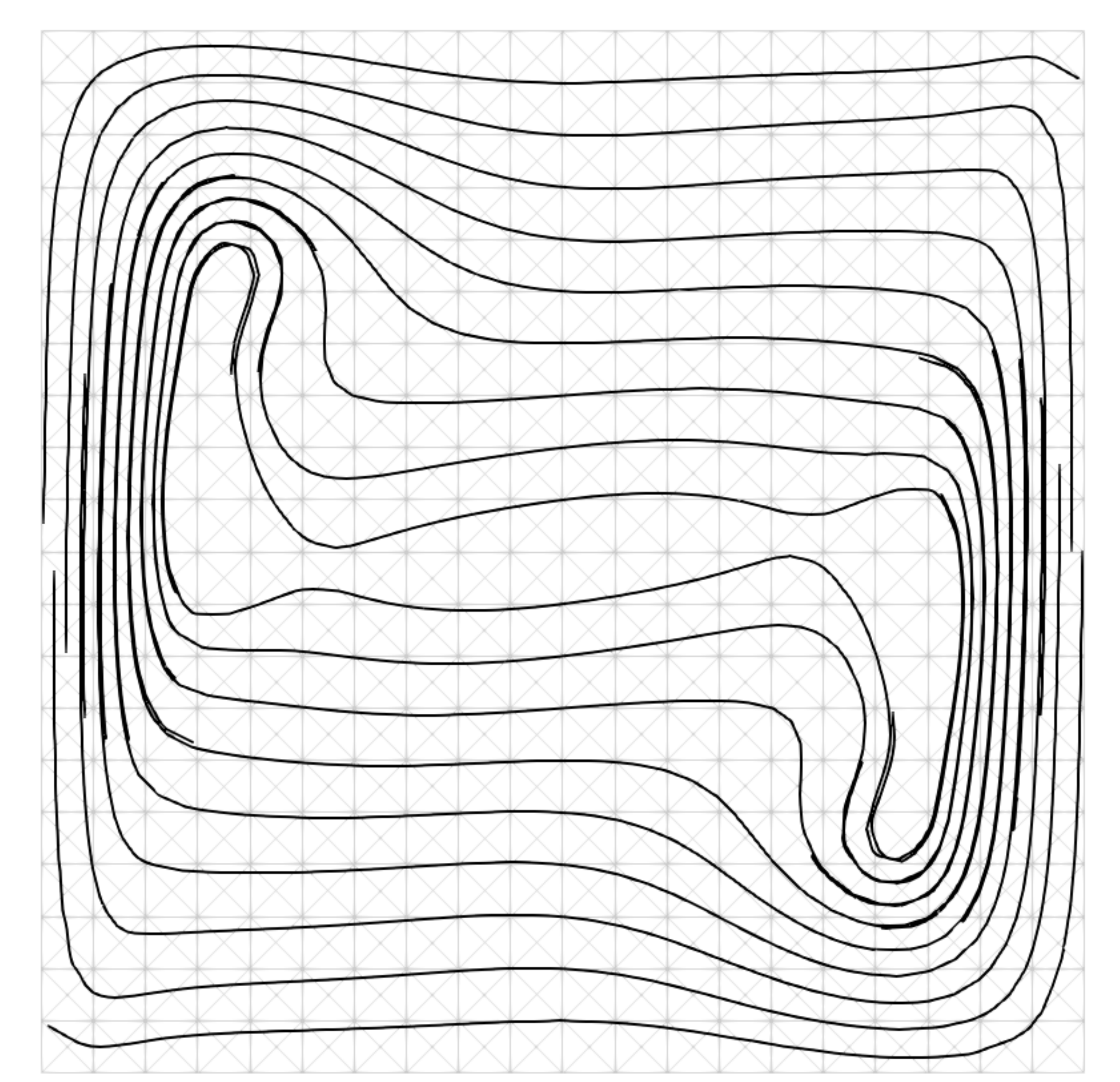}\quad
	\includegraphics[height=5.5cm]{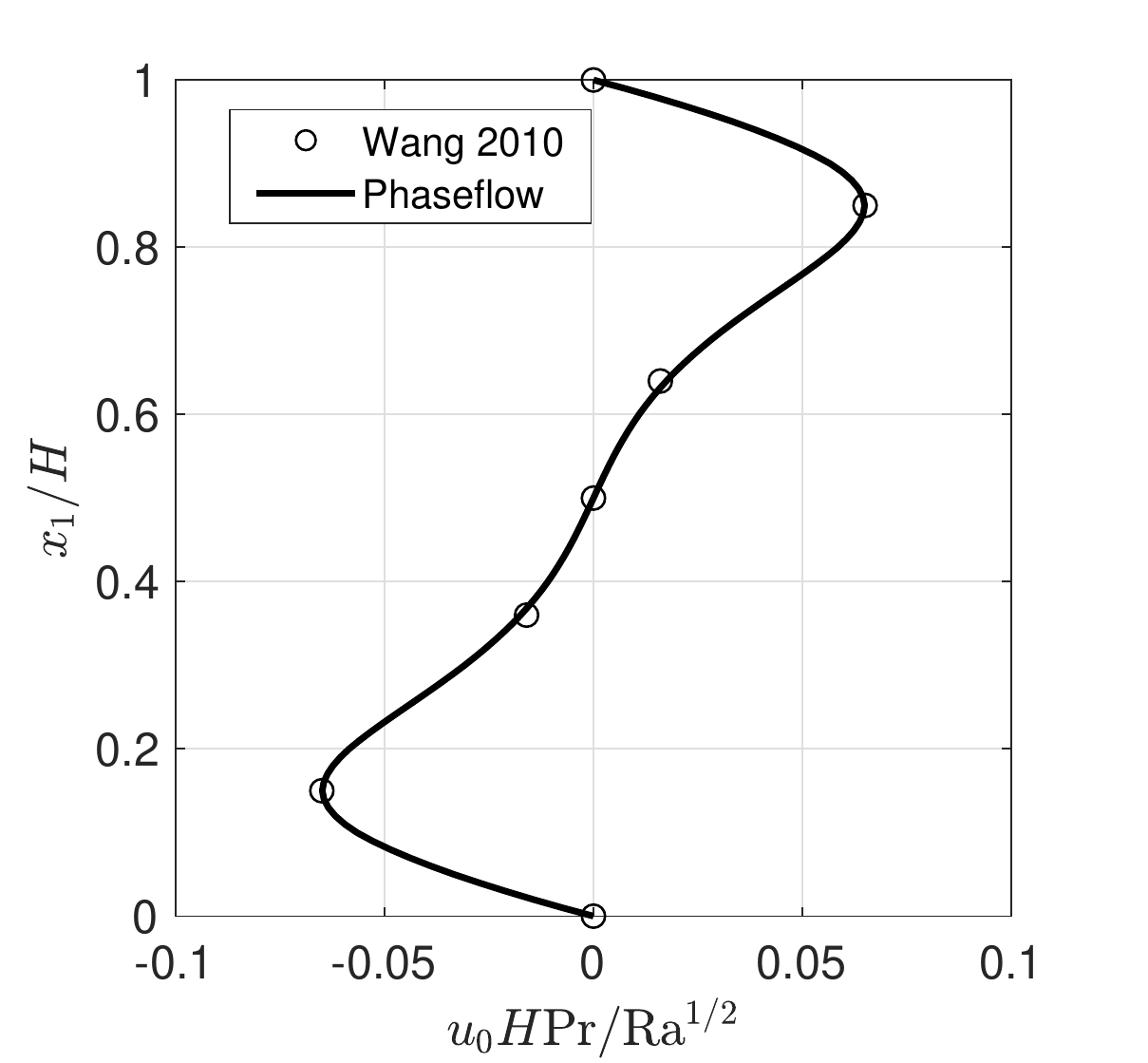}
	\caption{Left: Result from heat-driven cavity test ($\mathrm{Ra} = 10^6$).
		The mesh is shown in translucent gray.
		Velocity streamlines are shown in black.
		As expected, we see the flow circulating, being forced upward near the hot wall
		and downward near the cold wall. \newline
		Right: The horizontal component of velocity is plotted along the vertical centerline.
		The solution agrees very well with the data published in \cite{wang2010comprehensive}.}
	\label{fig:HeatDrivenCavity}
\end{center} \end{figure}

To verify the coupled energy equation without phase-change,
we compare to the heat-driven cavity benchmark data published in \cite{wang2010comprehensive}.
This problem fits our general model with 
constant viscosity  $\mu_S = \mu_L = 1$ and arbitrarily large Stefan number.
We handle buoyancy with an idealized linear
\footnote
{
	The method handles nonlinear bouyancy with (\ref{eq:GateauxDerivative}).
	Phaseflow's test suite includes a benchmark with the nonlinear density anomaly of water.
}
Boussinesq model
\begin{equation} \label{eq:Buoyancy}
	\mathbf{f}_B(T) = \frac{\mathrm{Ra}}{\mathrm{Pr}\mathrm{Re}^2}T
	\begin{pmatrix} 0 \\ -1 \end{pmatrix},
\end{equation}
The momentum equation (\ref{eq:momentum}) is scaled such that
the Reynolds number $Re$ is always unity.
For this benchmark, the Rayleigh number is $\mathrm{Ra} = 10^6$
and the Prandtl number is $\mathrm{Pr} = 0.71$.
This Rayleigh number is considered to be high, and demonstrates substantial natural convection.
We set homogeneous Dirichlet (i.e. no slip) boundary conditions on the velocity
$\mathbf{u}_D = \mathbf{0}$,
and non-homogeneous Dirichlet boundary conditions for the temperature
with hot and cool temperatures, $T_h = 0.5$ and $T_c = -0.5$, 
respectively on the left and right vertical walls, i.e.
\begin{equation} \label{eq:DifferentialTemperatureBC}
	T_D(\mathbf{x}) =
	\begin{cases}
		T_h & $for $ x = 0, \\
		T_c & $for $ x = 1
	\end{cases}
\end{equation}
Again we solve the unsteady problem until it reaches steady state.
For this we initialize the velocity field to zero (i.e. $\mathbf{u}_0 = \mathbf{0}$)
and the temperature field to vary linearly between
the hot and cold walls (i.e. $T_0(\mathbf{x}) = T_h + x(T_c - T_h)$).
Unlike with the lid-driven cavity, 
here we cannot obtain the steady state solution in a single time step,
because the initial guess is not sufficient
for the Newton method to converge.
Therefore we solve a sequence of time steps using $\mathrm{\Delta} t = 0.001$.
Figure \ref{fig:HeatDrivenCavity} shows the successful result with further discussion.

\subsection{The Stefan Problem} \label{section:StefanProblem}
To verify the energy equation with phase-change,
we compare to the analytical 1D Stefan problem as written in \cite{alexiades1992mathematical},
with parameters comparable to the octadecane PCM melting benchmark in Section \ref{section:2DMeltingPCM},
including the Stefan number $\mathrm{Ste} = 0.045$.
This problem fits our general model with 
nullified buoyancy $f_B = 0$,
unity Prandtl number,
and zero initial velocity $u_0 = 0$ (which remains zero).
Solving the Stefan problem as a special case
of the coupled problem results in an unnecessarily large system.
In this case, the vast majority of degrees of freedom are trivial.
From an implementation perspective, this exercise is quite valuable,
because we test the exact lines of code which are used for the coupled problem.

We again set temperature boundary conditions (\ref{eq:DifferentialTemperatureBC}),
this time with $T_h = 1$ and $T_c = -0.01$.
The initial temperature field is set such that a thin layer of melt exists
near the hot wall, with the rest of the domain at the cold wall temperature, i.e.
\begin{equation} \label{eq:MeltingInitialTemperature}
	T_0(x) = 
	\begin{cases}
		T_h & $for $ x \leq x^*_0, \\
		T_c & $otherwise $
	\end{cases}
\end{equation}
We parameterize this initial PCI position as 
\begin{equation} \label{eq:InitialPCIPosition}
x^*_0 = \frac{L}{N_0} 2^{1 - q}
\end{equation}
where $q$ is the number of initial hot wall refinement cycles.
This ensures that for the given initial mesh, the thinnest possible layer of melt exists.
For the semi-phase-field regularization (\ref{eq:phi}) we set parameters $T_r = 0, r = 0.01$.

\begin{figure}[tbp] \begin{center}
	\includegraphics[height=5.5cm]{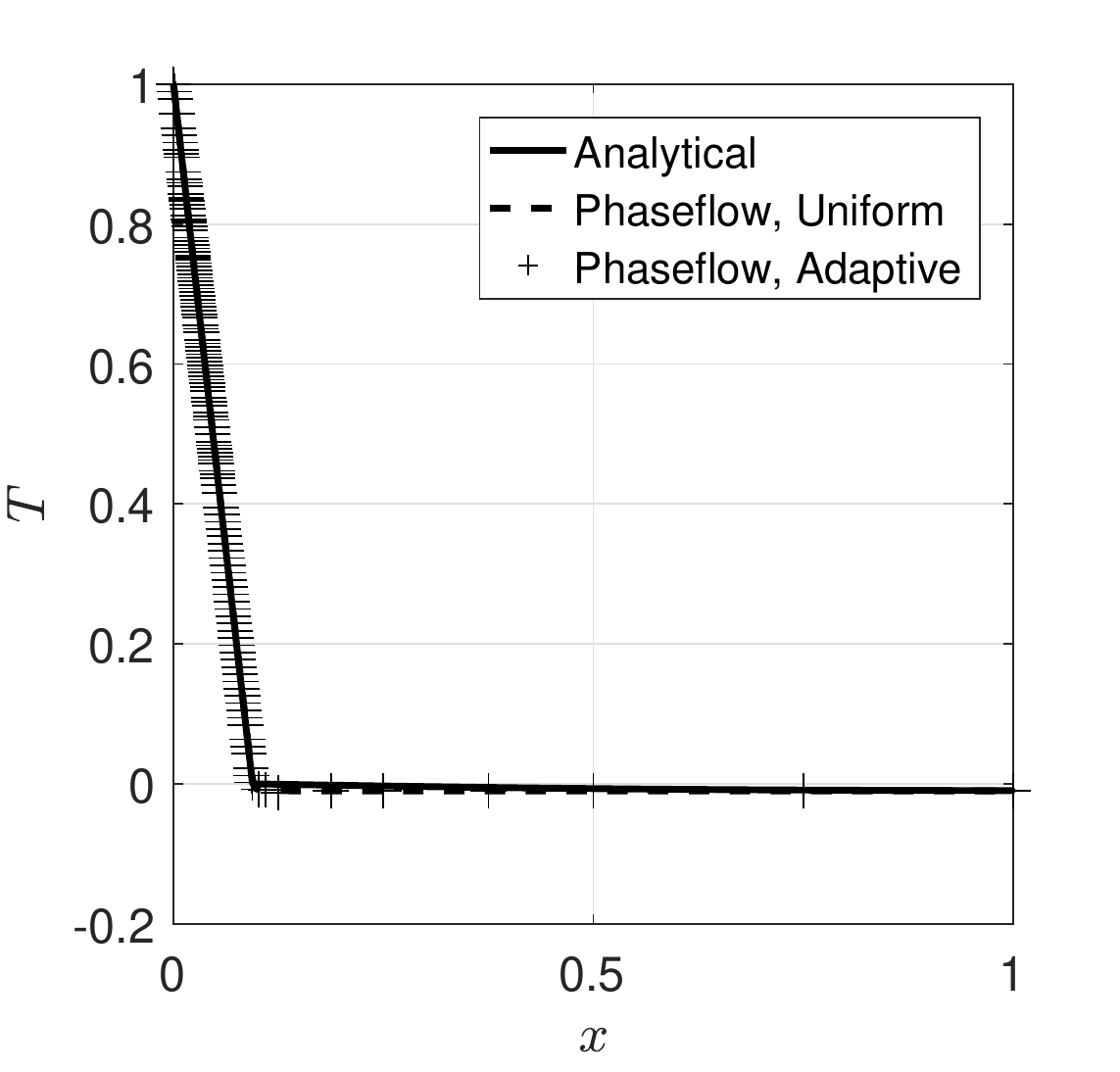}
	\includegraphics[height=5.5cm]{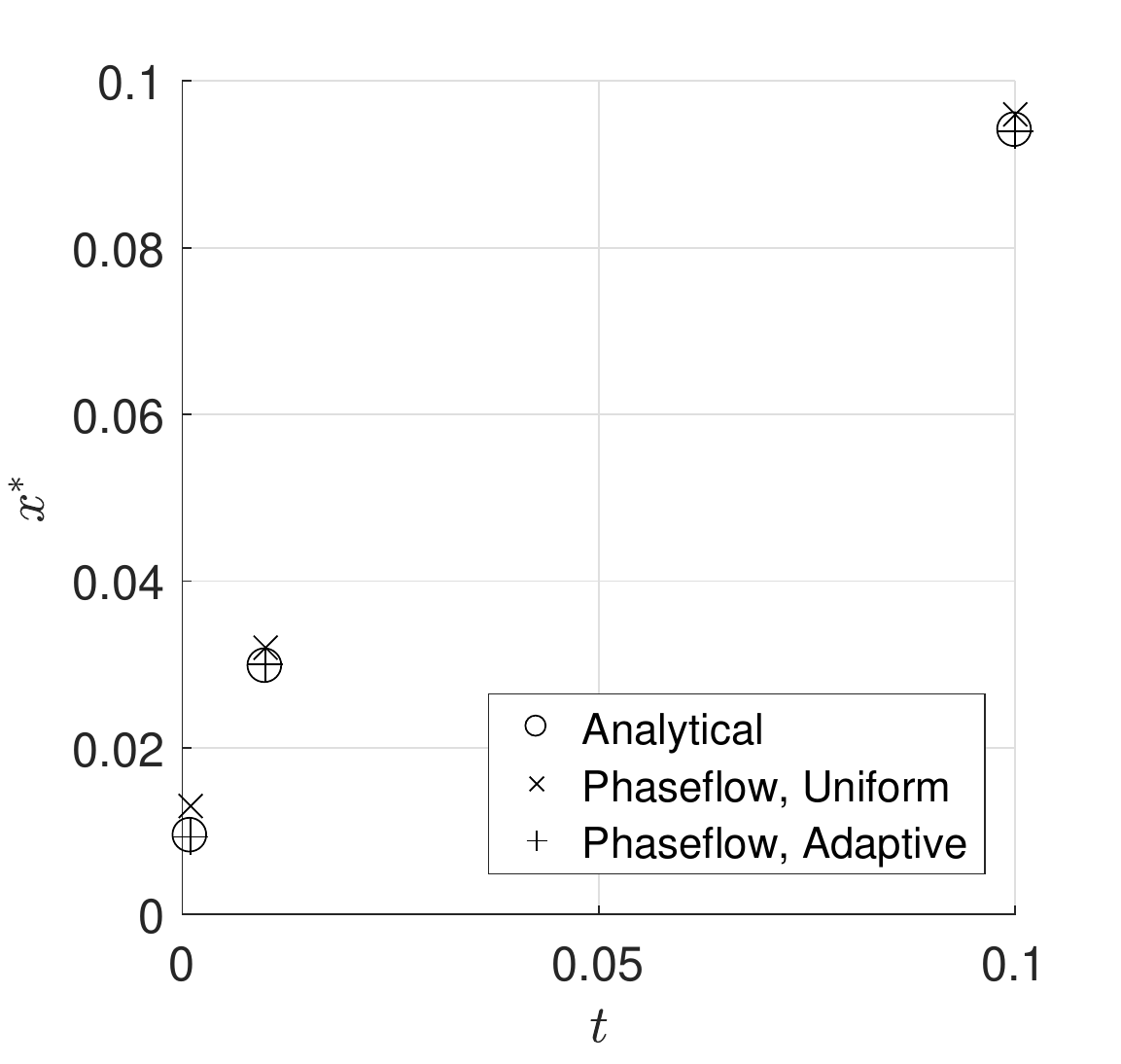}
	\caption[]
	{
		Left: Comparison at $t = 0.1$ between the analytical Stefan problem solution,
		Phaseflow using a uniform mesh of 311 cells\footnotemark,
		and Phaseflow using AMR (with 133 cells in the final adapted mesh).
		The AMR solution is shown with a marker at every mesh vertex.
		This emphasizes the much smaller cell sizes in the wake of the PCI.
		\newline
		Right:	Comparison of the PCI positions at three times between
		the analytical, uniform, and AMR solutions.
		We observe a bias in the uniform mesh solution which shrinks over time.
		This is primarily because of the mesh-dependent initial PCI position (\ref{eq:InitialPCIPosition}).
}
	\label{fig:StefanProblem}
\end{center} \end{figure}

We simulate until time $t_f = 0.1$,
using the time step size $\mathrm{\Delta t} = 0.001$.
In testing, this size was needed to bound the point-wise error between 
the numerical and analytical solutions from above 
by $T(x) - T_{\mathrm{exact}}(x) < 0.01 \quad \forall x \in \{0, 0.025, 0.05, 0.075, 0.1, 0.5, 1\}$ 
at time $t = t_f$ (shown in Figure \ref{fig:StefanProblem}).
This discrete set of verification points is a good proxy for the global solution.
Figure \ref{fig:StefanProblem} shows the successful result and further discussion.

We also take this opportunity to demonstrate AMR. We set the adaptive goal
(\ref{eq:AdaptiveGoalPhase})
with tolerance $\epsilon_M = 10^{-6}$.
During the time-dependent simulation,
new cells are only added, and never removed,
i.e. the mesh is never coarsened.
This is an unfortunate limitation of the current AMR algorithm.
We see that cells have been clustered near the PCI,
and these clusters remain everywhere the PCI has been during the time-dependent simulation.
Rightward of the PCI, the cells grow much larger,
e.g. the right half of the domain is covered by only two cells.
Even without coarsening,
already in 1D this is a large improvement.
The gains in 2D and 3D will be even greater.

\footnotetext
{
	For the uniform case, this is the minimum number of cells for which the Newton method would converge.
	Similarly for the AMR case, starting with a coarser initial mesh disrupts the Newton method.
}

\subsection{2D Convection-Coupled Melting of Octadecane PCM} \label{section:2DMeltingPCM}
To demonstrate the entire coupled system,
we present a preliminary result for the convection-coupled melting of the octadecane PCM benchmark
from \cite{wang2010comprehensive} and \cite{danaila2014newton}.
This problem uses all aspects of our general model (\ref{eq:VariationalForm}).
We set variable viscosity (\ref{eq:VariableViscosity}) (with $\mu_L = 1$ and $\mu_S = 10^8$), 
the buoyancy model (\ref{eq:Buoyancy})
(with $\mathrm{Ra} = 3.27 \times 10^5$, $\mathrm{Pr} = 56.2$, and $\mathrm{Re} = 1$),
no-slip velocity boundary conditions $\mathbf{u}_D = \mathbf{0}$,
and again the temperature boundary conditions (\ref{eq:DifferentialTemperatureBC}) with $T_h = 1$ and $T_c = -0.01$.
Again we initialize the temperature field with (\ref{eq:MeltingInitialTemperature}) such that the simulation
begins with a thin layer of melt,
and initialize a stationary velocity field $\mathbf{u}_0 = \mathbf{0}$.
For the semi-phase-field (\ref{eq:phi}), we set regularization parameters $T_r = 0.01$ and $r = 0.025$.
For the time-dependent simulation, we set $\mathrm{\Delta} t = 1$.
For AMR, we again set the adaptive goal (\ref{eq:AdaptiveGoalPhase}) from our Stefan problem example,
but with tolerance $\epsilon_M = 10^{-5}$ until $t = 36$ and $\epsilon_M = 0.5 \times 10^{-5}$ after that time.

\begin{figure}[tbp]
	\begin{center}
		\includegraphics[height=5.5cm]{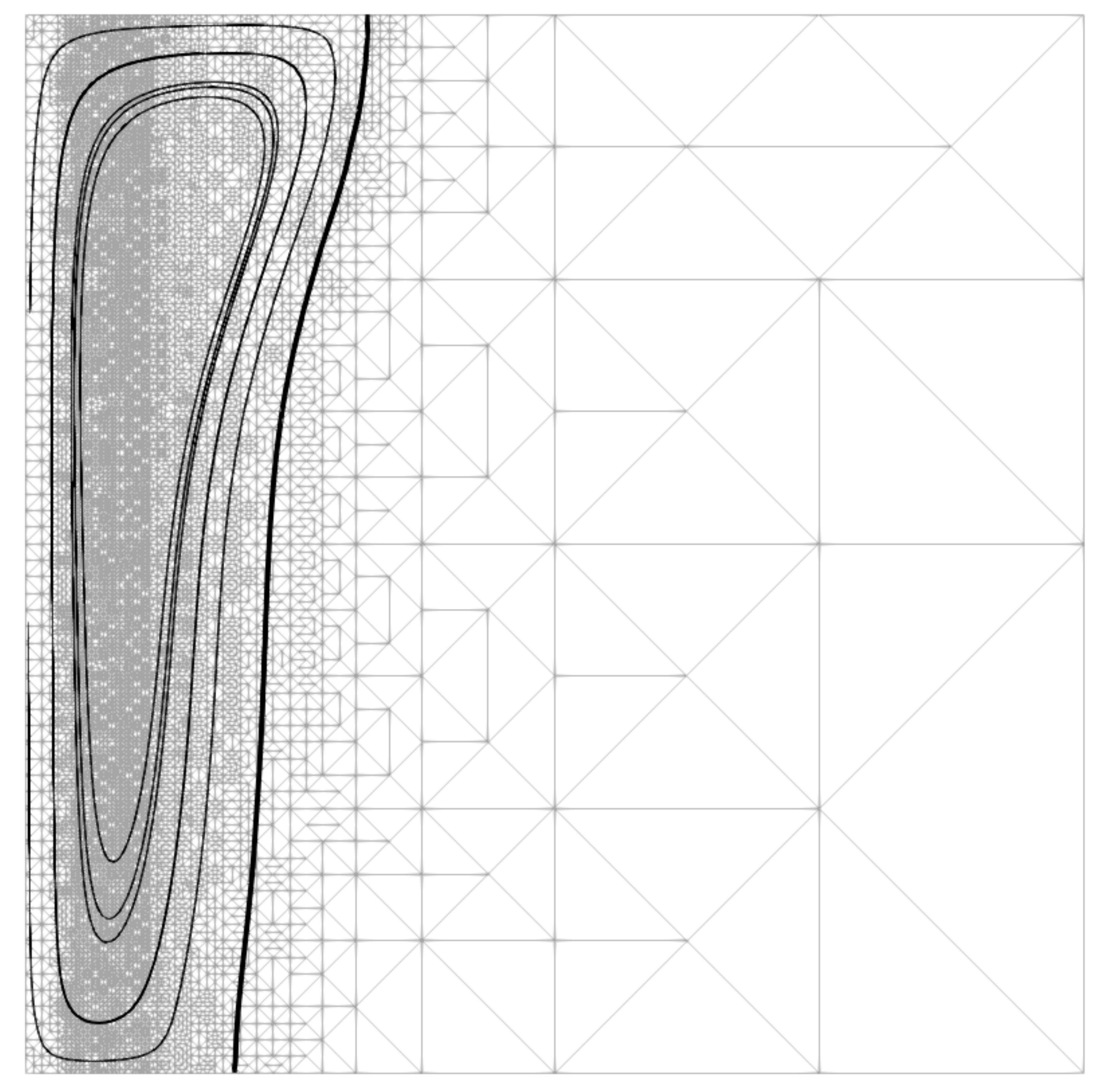}
		\includegraphics[height=5.5cm]{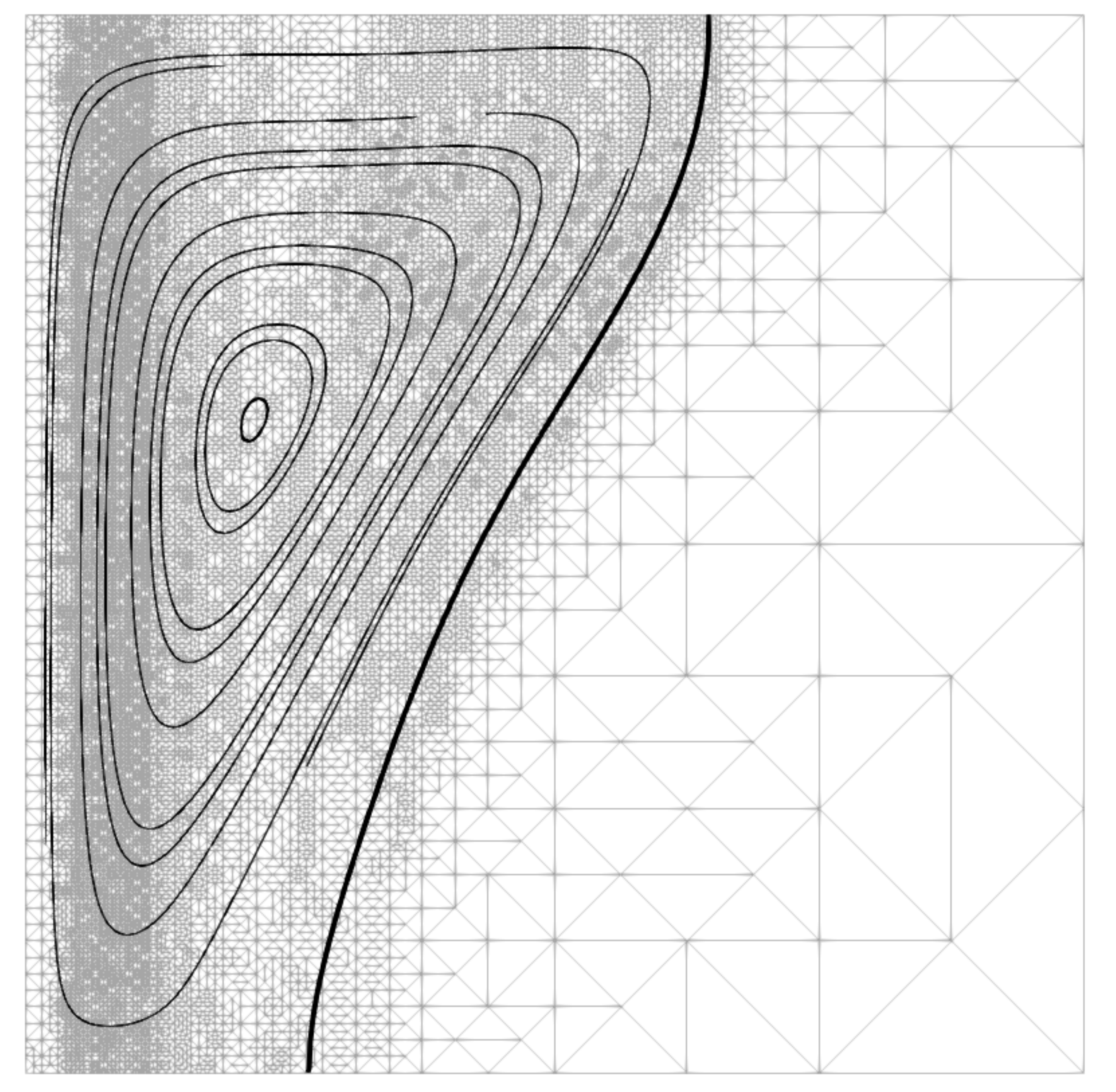}
		\caption[]
		{
			Preliminary result for the 2D convection-coupled octadecane PCM melting benchmark.
			Solutions are shown at simulated times $t = 36$ and $t = 80$ (on the left and right).
			The mesh is shown in translucent gray to highlight the adaptive mesh refinement (AMR).
			The PCI is shown as a thick black temperature isoline.
			Velocity streamlines are shown as thin black lines.
			This result looks promising when compared to
			octadecane PCM melting results from \cite{wang2010comprehensive} and \cite{danaila2014newton}.
			We see the flow circulating,
			and the PCI advanced more quickly at the top than at the bottom.
			We also see that the lack of grid coarsening is becoming expensive.
			By the final time, many refined cells from earlier times are likely not needed.
		}
		\label{fig:MeltingPCM}
\end{center} \end{figure}

The result in Figure \ref{fig:MeltingPCM} is promising;
but compared to the benchmark,
we see that the PCI has not advanced far enough by simulated time $t = 80$.
We still need to investigate the effects of our choices for
$\mathrm{\Delta} t$, $r$, $x^*_0$, $M(\mathbf{w})$, $\epsilon_M$, and the initial mesh refinement.
Of these, we know that a much smaller smoothing parameter $r$ was used in \cite{danaila2014newton}.

\subsection{3D Convection-Coupled Melting} \label{section:3DMeltingPCM}

\begin{figure}[tbp] \begin{center}
	\includegraphics[height=5.5cm]{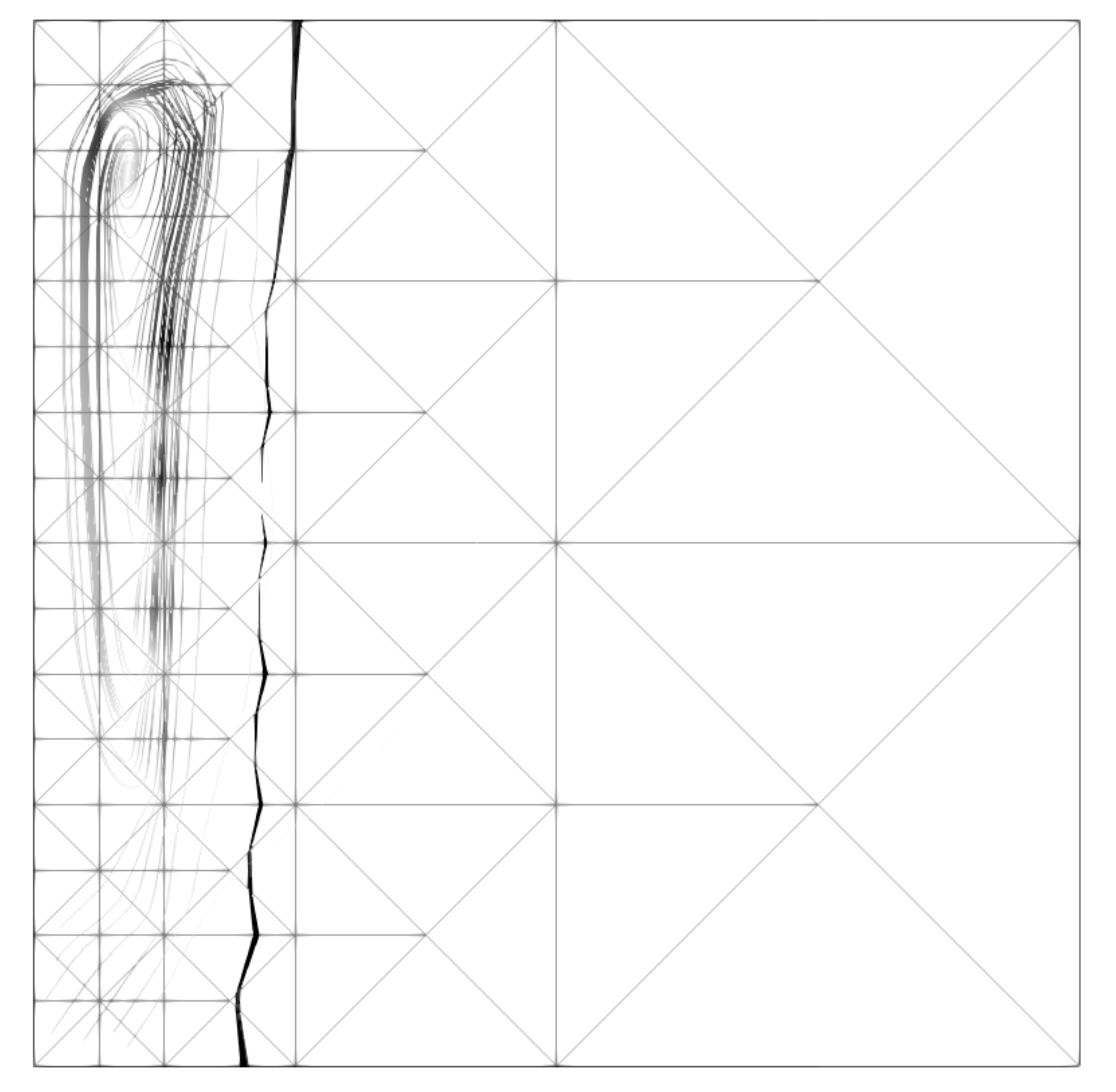}
	\includegraphics[height=5.5cm]{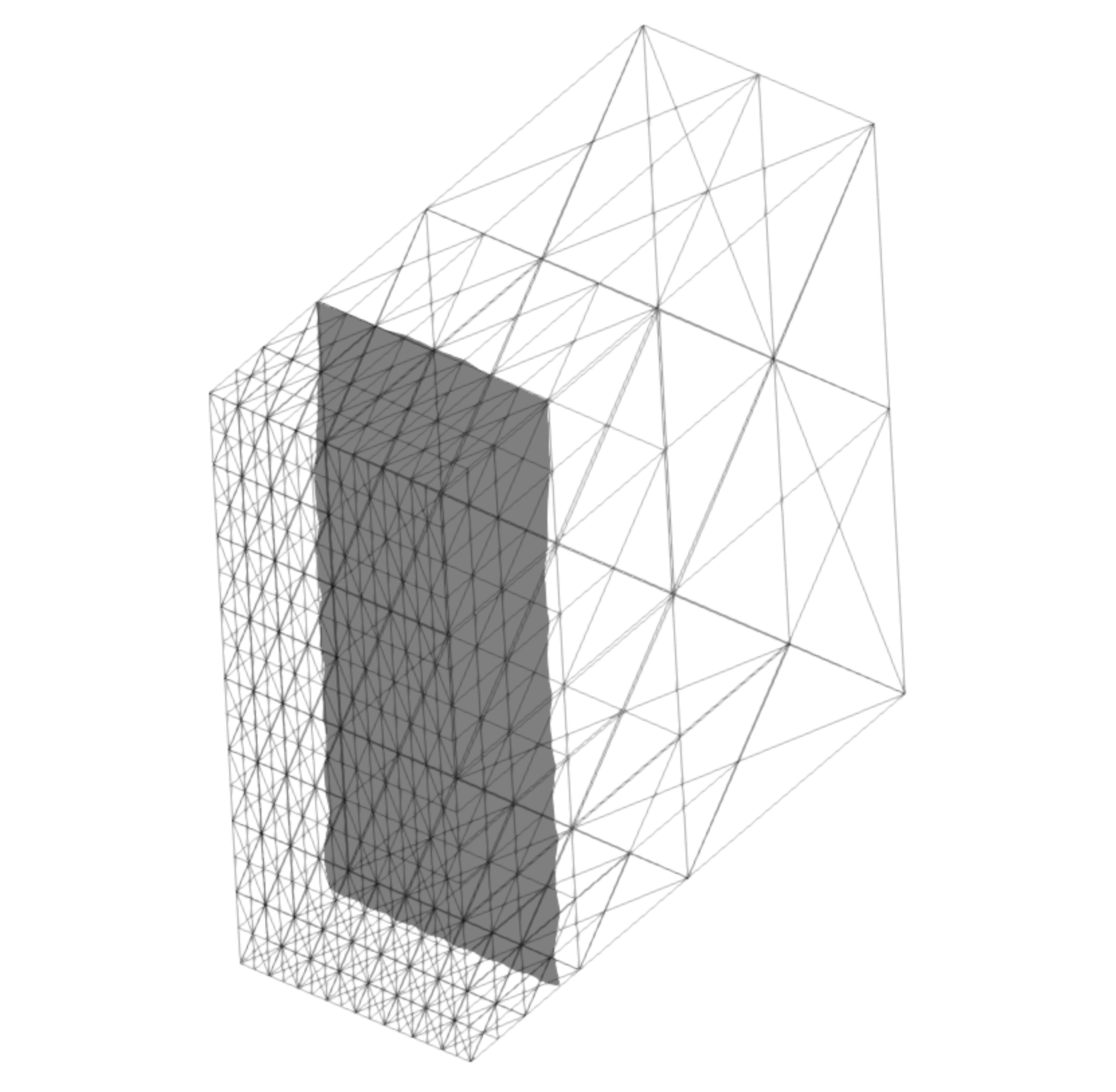}
	\caption
	{
		Preliminary result for 3D convection-coupled melting,
		with front and isometric views (on the left and right).
		The mesh is shown in translucent gray to highlight the local  mesh refinement.
		The PCI is a darker translucent gray iso-surface.
		The streamlines are colored by a grayscale from white to black as velocity magnitude increases.
		We use a coarse mesh and omit AMR,
		because the refined 3D problem quickly exceeded the capabilities of the desktop-scale computers used for this work.
		Despite the coarse mesh,
		we still observe the primary features we expect in the solution.
		The flow is circulating,
		and we see that the top of the PCI advancing more quickly than the bottom.
	}
	\label{fig:3DMeltingPCM}
\end{center} \end{figure}

The dimension-independent implementation,
facilitated by the abstractions from FEniCS,
allow us to quickly demonstrate a 3D example.
We consider the a problem similar to the 2D convection-coupled melting in Section \ref{section:2DMeltingPCM},
but with the domain extruded in the $z$ direction,
adiabatic no-slip boundary conditions on the walls parallel to the $z$ plane,
similarity parameters $\mathrm{Ste} = 1$, $\mathrm{Ra} = 10^6$, $\mathrm{Pr} = 0.71$,
and numerical parameters $\mu_S = 10^4$, $r = 0.05$. 
For the previous examples, we employed the "full" Newton method with $\omega = 1$.
For this problem, we relaxed the Newton method with a factor of $\omega = 0.8$.
Figure \ref{fig:3DMeltingPCM} shows the preliminary result.

\subsection{Convergence}

To verify the accuracies of the finite difference time discretization,
finite element space discretization,
and Newton linearization methods,
we consider the 1D Stefan problem
from Section \ref{section:StefanProblem}.
The mixed finite element formulation was shown
to be second order accurate for the incompressible Navier-Stokes equations
in \cite{brezzi1991mixed}.
We are not aware of such a result for the energy-coupled problem.
Here, as a first step, we focus only on the energy equation with phase-change.

\begin{figure}[tbp] \begin{center}
	\includegraphics[height=5.5cm]{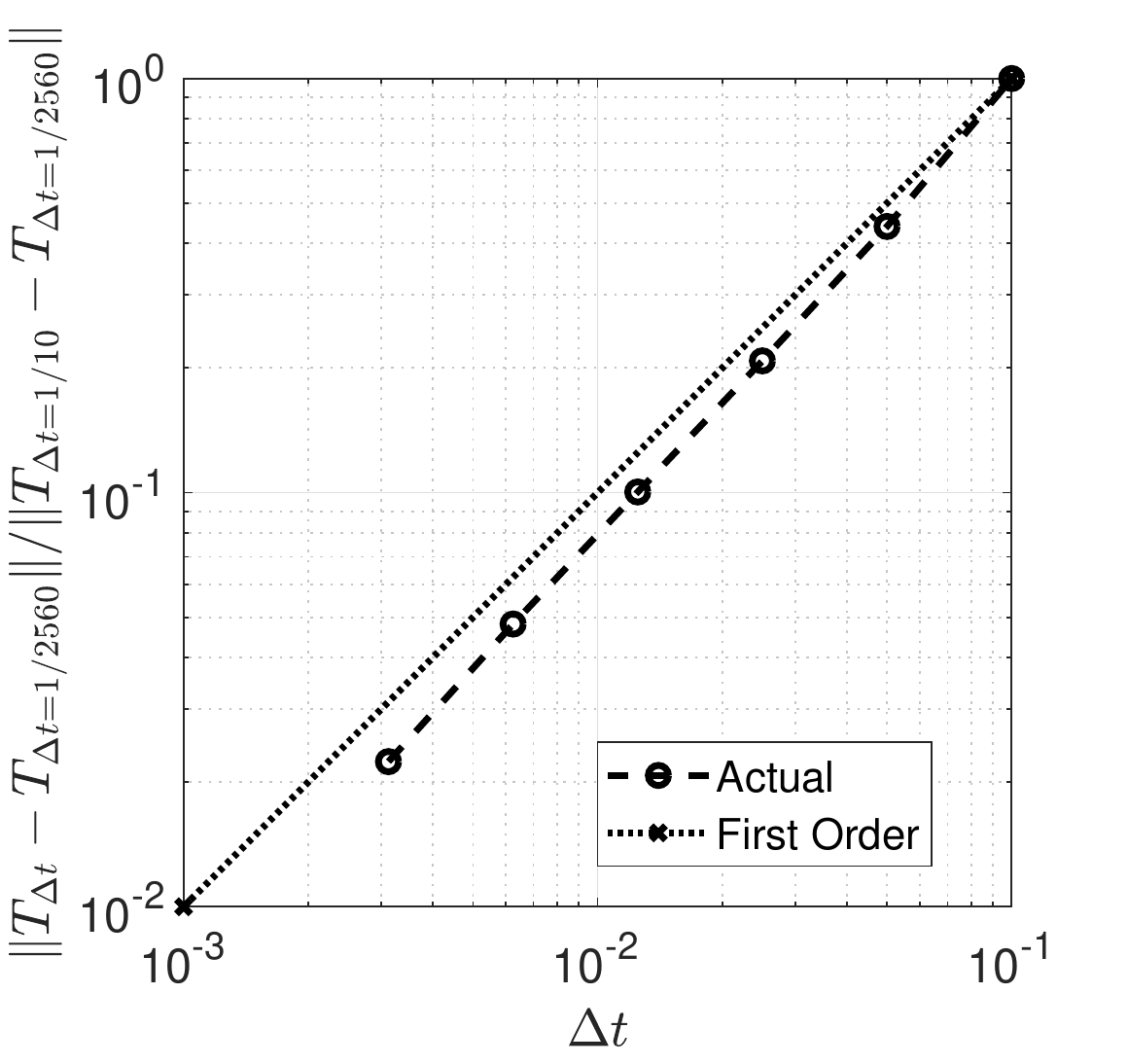}
	\includegraphics[height=5.5cm]{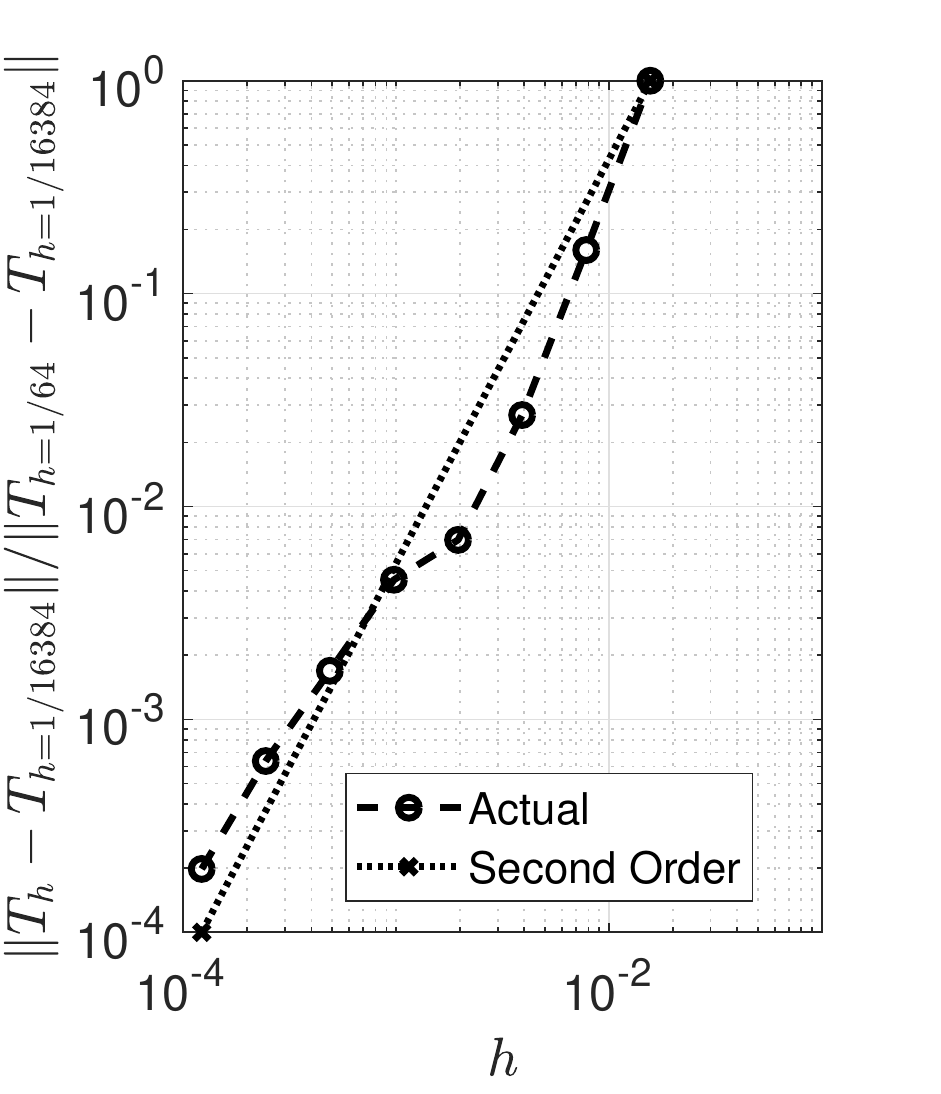}
	\includegraphics[width=0.49\linewidth]{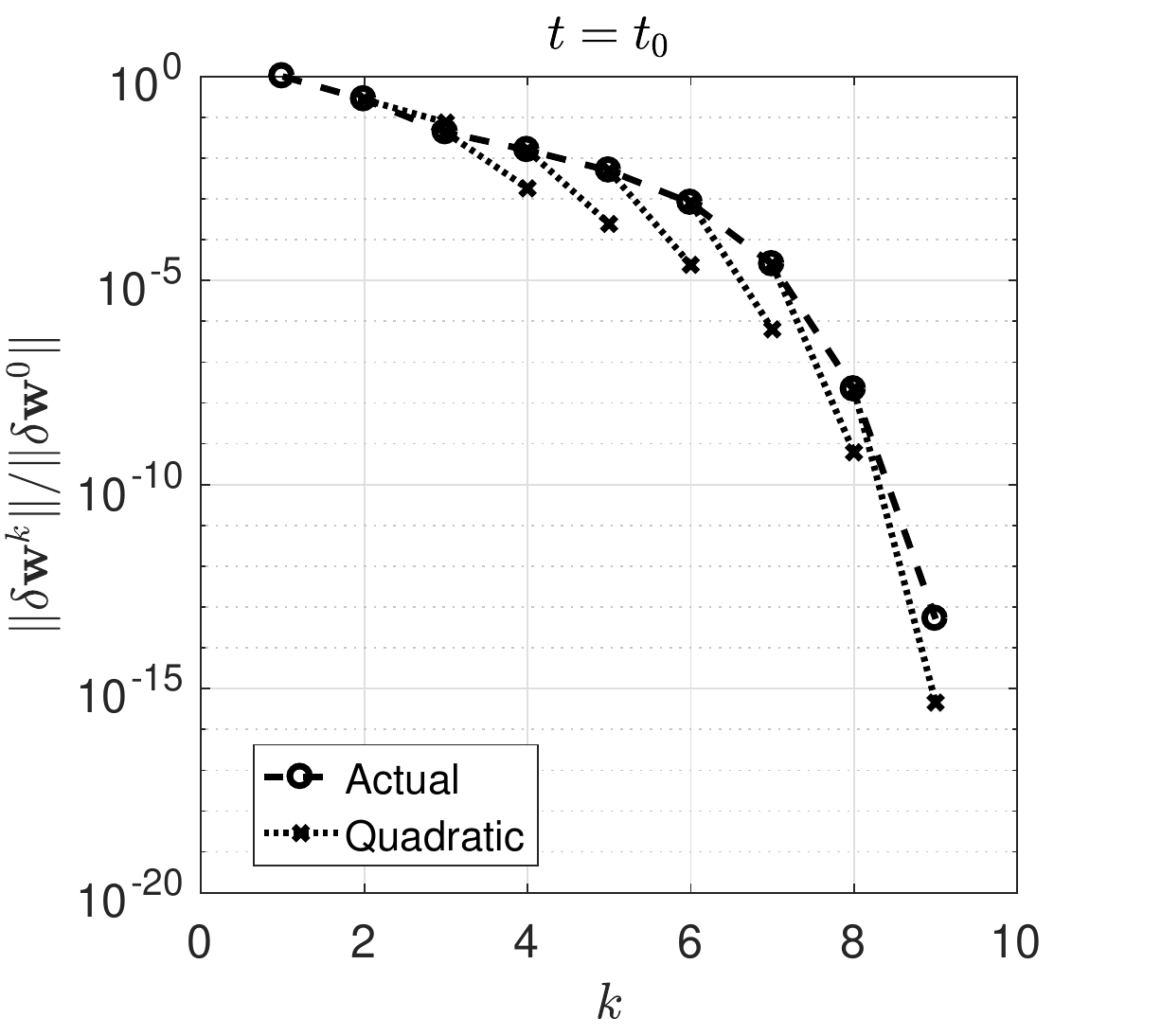}
	\includegraphics[width=0.49\linewidth]{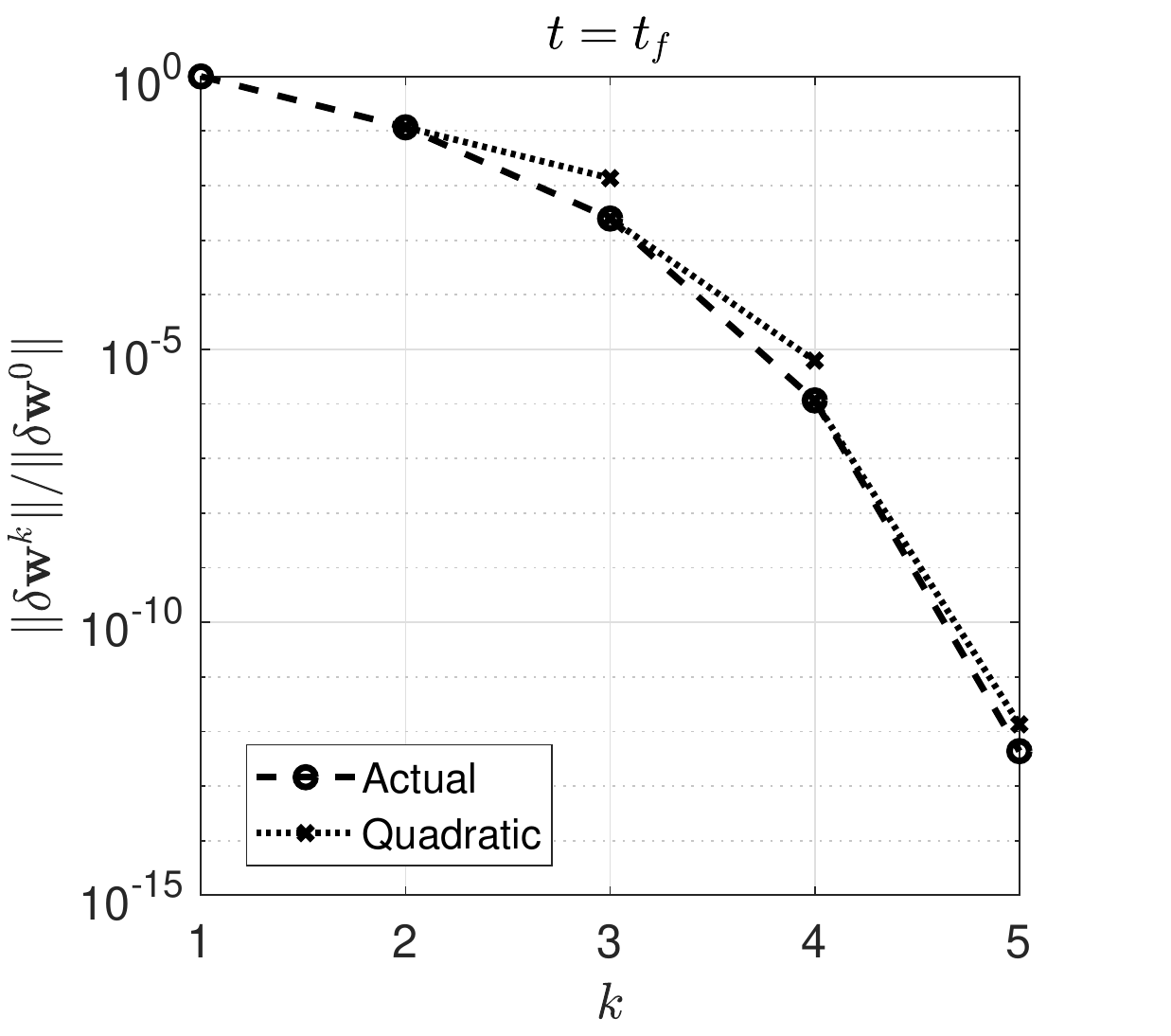}
	\caption
	{
		Convergence study results for
		Phaseflow applied to the Stefan problem from Section \ref{section:StefanProblem}.
		Top: Temporal (on the left) and spatial (on the right) convergence results
		based on the solution at the final time $t_f$.
		\newline
		Bottom: Newton method convergence shown for multiple time steps.
		Each sub-plot has a series of lines which represent ideal (quadratic) iterations.
	}
	\label{fig:Convergence}
\end{center} \end{figure}

Based on the choices of discretizations, 
with fully implicit Euler for time and finite elements for space,
we expect first order convergence in time and second order in space.
Figure \ref{fig:Convergence}
compares this with the actual convergence orders 
of Phaseflow's solution from Section \ref{section:StefanProblem}.
With respect to the time step sizes $\delta t$, 
the observed convergence order is only slightly higher than expected.
With respect to the grid spacing $h$,
for a sufficiently smooth solution
with well-behaved data, we should expect second order accuracy.
Though there is some deviation from second order near $h = 0.001$,
the results in Figure \ref{fig:Convergence} show good agreement.

Next, we verify the accuracy of the Newton method.
Figure \ref{fig:Convergence} shows its convergence behavior for multiple time steps,
highlighting the difference in performance between the first and last time steps.
Ideally, Newton's method will converge quadratically 
for well-posed problems with a suitable initial guess.
At earlier times, a few Newton iterations pass before reaching approximately quadratic convergence.
Later times converge slightly better than quadratically.

\section{Conclusions and Outlook}
In this work,
we presented the computational tool Phaseflow
for simulating the convection-coupled melting and solidification of PCM's.
Our work is based on the numerical method proposed in \cite{danaila2014newton},
and so we used an enthalpy formulated,
single-domain semi-phase-field,
variable viscosity,
finite element method,
with monolithic system coupling and global Newton linearization.
The primary difference between our numerical approach and that in \cite{danaila2014newton}
is the mesh adaptivity algorithm, 
where we employ the dual-weighted residual method for goal-oriented AMR.
We implemented the method into our open source Python module Phaseflow,
based on the finite element software library FEniCS,
and verified this against a series of classical benchmarks.
We obtained a promising result for the octadecane PCM convection-coupled melting benchmark.
Leveraging our dimension-independent implementation,
we contributed detailed convergence results for the 1D phase-change problem,
and have applied the method to a preliminary 3D convection-coupled melting example.
Furthermore, we openly shared a Docker container which allows anyone 
to reproduce our results in the same software environment.
It is our hope that this facilitates the further development of this and related methods.
The method appears promising,
and Phaseflow is ready for application to interesting problems.

FEniCS was a good choice as the base of our implementation,
allowing us to focus on the models and numerical methods 
rather than on implementing standard algorithms.
Furthermore, the existing Docker software container
allowed us to quickly leverage the library.
Some difficulties do remain, primarily:
1. FEniCS lacks mesh coarsening capability, prohibiting the efficient application of AMR to the moving PCI problem.
2. The adaptive solver (using dual-weighted residual goal-oriented AMR)
has not been implemented for distributed memory systems,
therefore currently limiting the implementation's applicability to a single compute node,
prohibiting realistic 3D applications.
On the other hand, there is yet potential to simplify Phaseflow with existing FEniCS features.
Most interestingly, there is an automatic differentiation capability 
which could serve as an alternative to computing the Gâteaux derivative (\ref{eq:GateauxDerivative}) directly.
During the development of Phaseflow,
we have successfully applied this feature to some simplified cases;
and we would like to further explore this capability.

Next steps of our work include the development of robustness features, including...
1. \textit{a priori} bounds on the time step size $\mathrm{\Delta} t$,
based on the properties of the linear system (\ref{eq:LinearizedProblem}) and initial values (\ref{eq:InitialValues}),
2. an algorithm for obtaining the initial mesh on which to obtain the first solution to begin AMR.
The latter is important, because we discovered a limitation with relying on AMR to resolve strong nonlinearities.
Typically, goal-oriented AMR is used to provide either 
the most accurate solution (with respect to the goal) for a given cost,
or the minimum-cost discretization for a given accuracy (with respect to the goal).
In the case of the presented method,
AMR is required for the Newton method to converge 
and to hence obtain the first solution in the hierarchy.
Handling these issues is necessary before practically applying this method to a wider range of realistic problems.

\bibliography{zimmerman_kowalski_2017}{}
\bibliographystyle{plain}

\ifx\justbeingincluded\undefined
\end{document}
\fi